\documentclass[
reprint
,aps
,amssymb
,amsmath
,superscriptaddress
]{revtex4-1}

\usepackage{graphicx}
\usepackage{bbold}
\usepackage{dcolumn}
\usepackage{fancyref}
\usepackage{bm}
\usepackage{color}
\usepackage[toc,page]{appendix}
\usepackage[colorlinks=true,allcolors=blue]{hyperref}
\usepackage[normalem]{ulem}
\usepackage[fleqn]{mathtools}

\newcommand{\vvec}[1]{\mathbf{#1}}
\newcommand{\ttens}[1]{\mathbf{#1}}

\def\mechScalar{m} 
\def\mechScalarInit{\mechScalar_{0}}
\def\strain{\varepsilon} 
\def\strainScalar{\varepsilon} 
\def\mechStrain{\strain_m}
\def\nonMechStrainOne{\strain_1}
\def\nonMechStrainTwo{\strain_2}
\def\unsymStrain{S}
\def\metric{g}

\def\basisTens{\hat{e}}
\def\basisTensMech{\ttens{\basisTens}_m}
\def\basisTensOne{\ttens{\basisTens}_1}
\def\basisTensTwo{\ttens{\basisTens}_2}

\newcommand{\latticeVec}[1]{\vvec{l}^{(#1)}}
\newcommand{\latticeVecInit}[1]{\vvec{l}^{(#1)}_{0}}
\def\latticeVecOne{\vvec{l}^{(1)}}
\def\latticeVecTwo{\vvec{l}^{(2)}}
\def\latticeVecOneInit{\vvec{l}_0^{(1)}}
\def\latticeVecTwoInit{\vvec{l}_0^{(2)}}

\def\defTens{\Lambda}
\def\rotationMat{\mathcal{R}}
\def\coarseRot{\phi}
\def\rightStretch{U}
\def\rightStretchMech{A}
\def\specialCompatibilityMatrix{B}

\def\principal{\lambda}   
\def\principalOne{\principal_1}
\def\principalTwo{\principal_2}

\def\ident{\mathbb{1}}

\def\stress{\sigma} 
\def\mechStress{\stress_m}
\def\nonMechStressOne{\stress_1}
\def\nonMechStressTwo{\stress_2}

\def\gamgam{\gamma} 
\def\poisson{\nu}   

\def\aA{A} 

\def\levi{\epsilon} 

\def\ww{w} 

\def\softMode{f}
\def\softModeOne{\softMode_1} 
\def\softModeTwo{\softMode_2} 

\def\stiffMode{g}
\def\stiffModeOne{\stiffMode_1} 
\def\stiffModeTwo{\stiffMode_2} 

\def\posr{\mathbf{R}} 
\def\posrFinal{\mathbf{r}} 
\def\dee{\mathrm{d}} 

\def\designAngle{\psi} 
\def\mechAngle{\theta} 
\def\mechAngleInit{\mechAngle_{0}}

\def\stiff{G} 

\def\interfaceAngle{\phi}

\begin{document}

\title{Duality and Sheared Analytic Response in Mechanism-Based Metamaterials}


\begin{abstract}

Mechanical metamaterials designed around a zero-energy pathway of deformation, known as a mechanism, have repeatedly challenged the conventional picture of elasticity.
However, the complex spatial deformations these structures are able to support beyond the uniform mechanism remain largely uncharted. 
Here we present a unified theoretical framework, showing that the presence of any uniform mechanism in a two-dimensional structure fundamentally changes its elastic response by admitting a family of non-uniform zero-energy deformations.
Our formalism reveals a mathematical duality between these stress-free strains, which we term ``sheared analytic modes'' and the supported spatial profiles of stress.
These modes undergo a transition from bulk periodic response to evanescent surface response as the Poisson's ratio $\poisson$ of the mechanism is tuned through an exceptional point at $\poisson=0$. 
We suggest a first application of these unusual response properties as a switchable signal amplifier and filter for use in mechanical circuitry and computation.

\end{abstract}

\author{Michael Czajkowski}
\affiliation{School of Physics, Georgia Institute of Technology, Atlanta, Georgia 30332, USA}
\author{D. Zeb Rocklin*}
\affiliation{School of Physics, Georgia Institute of Technology, Atlanta, Georgia 30332, USA}
\date{\today}

\maketitle

Classical elasticity is a field theory describing a structure's deformation from a single zero-energy shape. In contrast,  a growing number of strategies and methods to program special energy-free pathways of deformation directly into designer materials~\cite{Kadic2012,Bertoldi2017, Li2019, Dieleman2019, Liu2021, Meloni2021} generate structures with continuous manifolds of (nearly) zero-energy shapes.
Such pathways, known as mechanisms, fundamentally challenge the classical picture of elasticity by setting common elastic constants to anomalous zero or negative values~\cite{ Nicolaou2012, Kadic2012, Buckmann2014, Kane2014, Paulose2015, Paulose2015-2, Dudek2016, Wu2016,  Rocklin2017, Bertoldi2017, Fruchart2020, Bossart2021}. Still further structures have been able to generate previously forbidden (odd) elastic constants~\cite{Scheibner2020}, gyroscopic elastic behavior~\cite{Nash2015}, multistable~\cite{Jin2020} and hierarchical structures~\cite{Gatt2015, Coulais2018-2, Michel2019} as well as frustrated structures lacking a zero-energy elastic reference state~\cite{Kang2014, Armon2014}; and yet the generic consequences of a single mechanism on elastic response have still remained largely unexplored.

\begin{figure}[t]  
\begin{center}
    \includegraphics[width=0.48\textwidth]{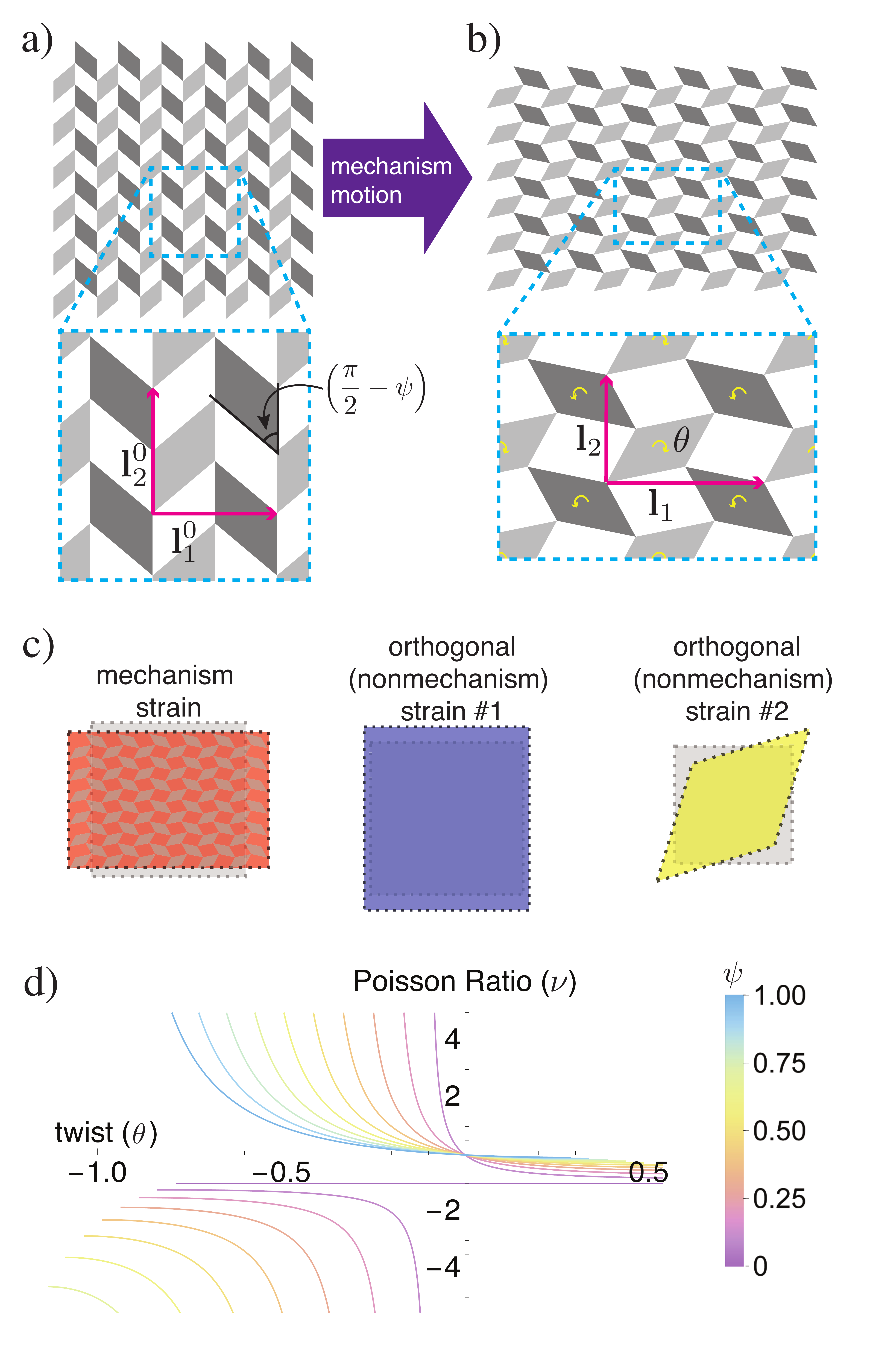}
    \caption{ \textbf{A characteristic class of planar mechanisms.} (\textbf{a}) The parallelogram-based mechanism designs we investigate are generated by setting a design angle $\designAngle$, and includes the canonical ``Rotating Squares'' pattern at the point $\designAngle=0$. (\textbf{b})  The mechanism itself is traversed via rotating each rigid parallelogram (dark grey) opposite to its neighbors (light grey), which alters the macroscopic strain as reflected in the lattice vectors (pink arrows). Light and dark grey block coloration is purely for ease of viewing. (\textbf{c}) Such a mechanism strain may be used to generate an orthonormal basis for strain which divides any arbitrary strain into mechanism and nonmechanism components. (\textbf{d}) The strains generated by varying the counter-rotation $\mechAngle$ for a variety of different $\designAngle$ (different lines) capture an arbitrary variety of Poisson's ratios.  \label{fig:main1} }
    \end{center}
\end{figure}

A recent series of investigations has revealed that a purely dilational mechanism fundamentally changes the response of a continuum material by introducing an associated space of conformal soft modes~\cite{Sun2012, Czajkowski2022, zheng2021continuum}. In the continuum limit of a perfect mechanism, these modes cost zero energy, and it was suggested that such a nontrivial soft mode space would come paired with any generic mechanism even outside the dilational limit~\cite{Rocklin2017, Czajkowski2021-2, zheng2021continuum, Czajkowski2022}, and some indicative nonlinear examples of this phenomenon have since been identified~\cite{zheng2021continuum}. 
However, by the introduction of a space of stress-free continuum responses, these mechanisms must also necessarily be eliminating stress-bearing response patterns that might have been supported in these systems. Therefore, the question presses: how is the overall space of supported deformations necessarily changed by mechanism design?

Here, rather than focus on a particular microstructure or type of mechanism, we explore how the presence of an arbitrary mechanism necessarily determines a two-dimensional structure's  non-uniform equilibrium response, in much the same way that conventional translational and rotational symmetries necessarily give rise to elastic waves as Goldstone bosons~\cite{chaikin2000principles}. We find that the impact of an arbitrary mechanism is to  generate spatial patterns of stress-free strains that are dual to the system's permitted stresses. Both are analytic in a particular set of sheared coordinates, which we introduce here, in which the shearing passes from a real transformation for mechanisms with conventional positive Poisson ratios ($\poisson >0$) to complex coordinates for auxetic mechanisms as in the purely dilational case~\cite{Czajkowski2022}.

In Sec.~\ref{sec:mechElasticity} we establish the alteration which a mechanism produces on the elastic energy, introducing appropriate new strain variables in a generalization of Voigt notation. In Sec.~\ref{sec:shearedAnalytics} we explore the impact of this energy, identifying a closed-form solution for the deformation patterns which are stress-free. We show that these deformations, which we term ``sheared analytic modes'', obey a duality with the force-balanced patterns of stress. These spaces of modes are completely determined by the Poisson's ratio and principal axes of the mechanism, and in Sec.~\ref{sec:spatialModes}, we show the existence of an exceptional point in the Poisson's ratio separating categories of spatial response in these systems. We illustrate the qualitative division between auxetic and anauxetic mechanism response using two illustrative analytic examples, further using the latter, a long strip geometry, to indicate a possible new device for mechanical computing.

\begin{figure*}[t]  
\begin{center}
    \includegraphics[width=0.99\textwidth]{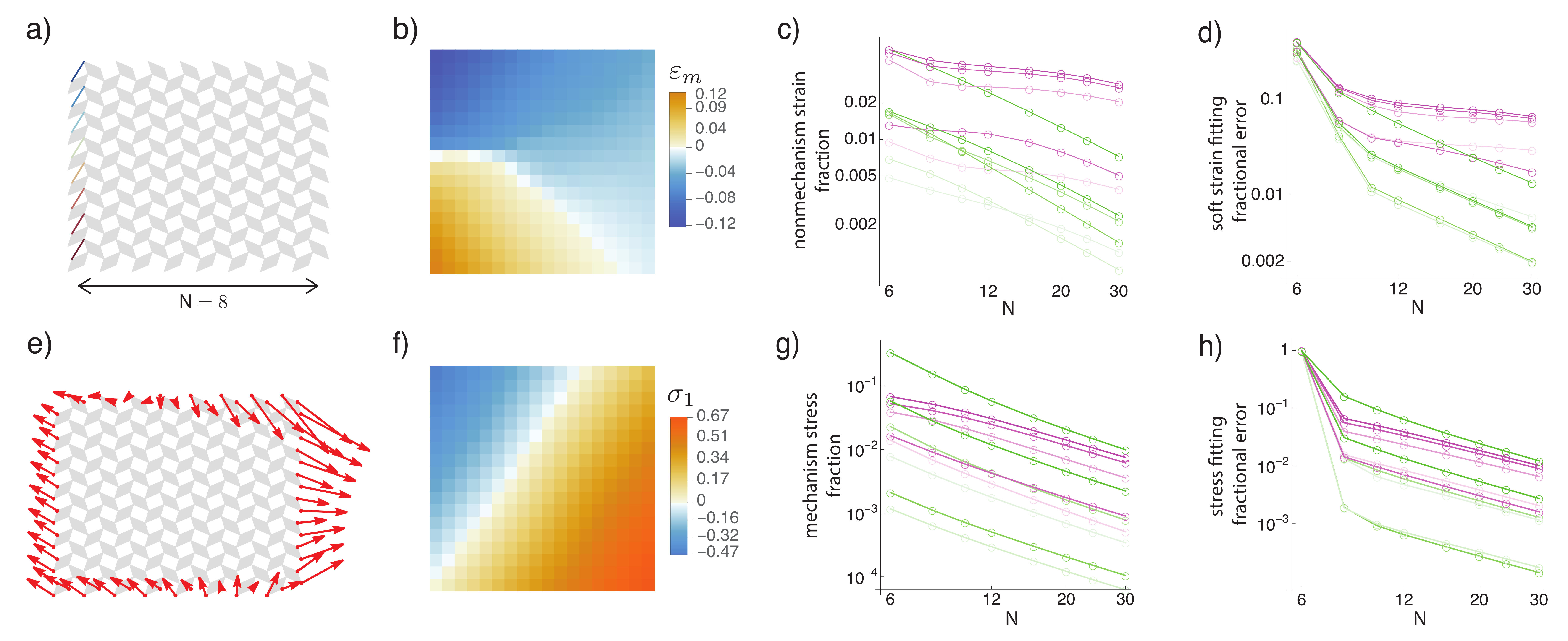}
    \caption{ \textbf{Unimode mechanics in lattice metamaterials}  (\textbf{a - d}) Minimal loading which is compatible with soft (stress-free) motions yields mechanism dominated soft strain patterns. (\textbf{a}) The systems deform due to a set of additional springs at the system left boundary which are constrained to extend according to a smoothly varying function through space. (\textbf{b}) The mechanism strain $\strain_m$ exhibits significant variation through space. (\textbf{c, d}) This mechanism strain dominates over the nonmechanism strain to an increasing degree, and the fitting to the global analytic form for soft strains improves as the lattice structure becomes finer. (\textbf{e-h}) More strict loading which is not compatible with a soft motion may still be deciphered using sheared analytic functions. (\textbf{e}) Loading is generated using a known force balanced mode, generated from a sheared analytic function, along with known nonaffine relaxation, applied to all boundary nodes and then numerically relaxed in the interior. (\textbf{f}) The nonmechanism stress is finite and varies through space. (\textbf{g, h}) This nonmechanism stress dominates increasingly, and the fitting to a sheared analytic mode improves as the continuum limit is approached just as with the soft strains.  \label{fig:main2} }
    \end{center}
\end{figure*}

\begin{figure*}[t]  
\begin{center}
    \includegraphics[width=0.99\textwidth]{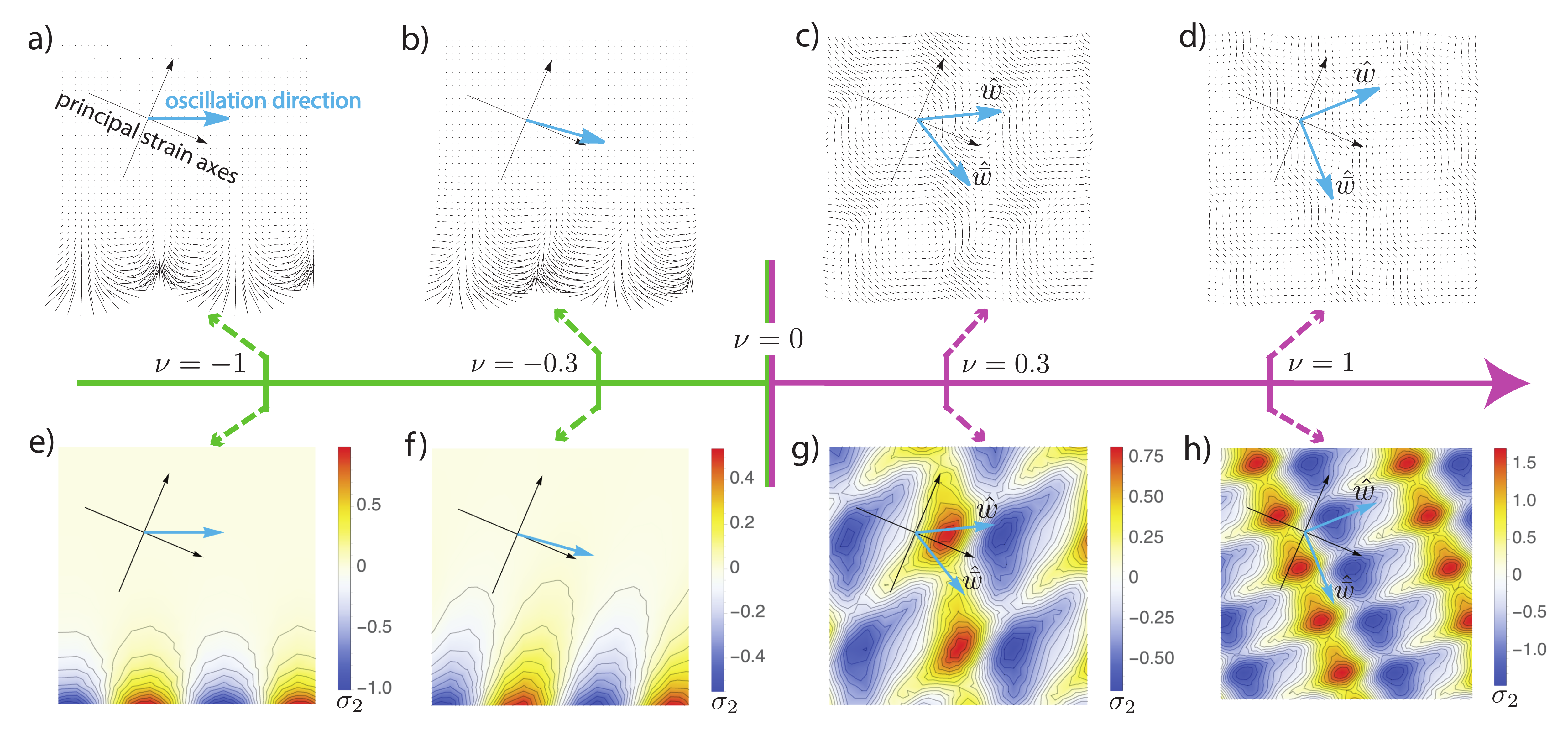}
    \caption{ \textbf{Spatial distribution of static unimode response near an open boundary}  (\textbf{a - d}) Stress-free continuum deformations of the unimode metamaterial in the semi-infinite plane, with arbitrarily oriented mechanism principal strain axes (large black arrows) will change character as an exceptional point is crossed at Poisson's ratio $\nu=0$. (\textbf{a}) For the maximally auxetic system ($\nu=-1$), an oscillatory set of displacements (black lines) along the open boundary (bottom edge) will oscillate paralell to the boundary (purple arrow) while decaying into the bulk along the perpendicular. (\textbf{b}) For a more generic auxetic system, bulk oscillation will be determined by a special direction (purple arrow) set by the specific boundary orientation and by $\nu$.   (\textbf{c, d}) On the other side of the exceptional point, the soft displacements of the anauxetic mechanism will oscillate into the bulk without decay, and the boundary conditions on the transverse waves may change with the addition of more modes when the longitudinal component is fixed. (\textbf{e-h}) The same unimode metamaterial, subjected to stressed boundary conditions will display identical spatial patterns in the stress distribution due to the duality.  \label{fig:main3} }
    \end{center}
\end{figure*}


\section{Elasticity theory for generic planar mechanism metamaterials \label{sec:mechElasticity}}

Consider an elastic solid undergoing a deformation such that matter initially located at material coordinates $\posr = (x,y)$ is displaced by  $\vvec{u}(\posr)$.
Because the system is translationally invariant, the energy depends on displacement gradients, rather than the bare displacements. In addition, because it is rotationally invariant, the energy depends only on the symmetrization of these gradients, the small strains
$\strain_{ij} \equiv (\partial_i u_j + \partial_j u_i)/2$~\cite{landau1986theory, chaikin2000principles}. The local energy then takes the general and well-known form $C_{ijkl} \strain_{ij} \strain_{kl}/2$, in terms of the three strain components  $\strain_{11}, \strain_{12}, \strain_{22}$.
 
In contrast, consider an elastic structure containing a \emph{mechanism}. In a coarse description, such a system is defined to contain a particular strain pathway that, as was the case with rotations, does not contribute to the elastic energy density. As shown in Appendix~\ref{app:generalizedVoigt}, it is still always possible to construct orthonormal components of strain which separate this \emph{mechanism strain} $\mechStrain$ from the non-mechanism strains $\nonMechStrainOne, \nonMechStrainTwo$. Such variables inherently span all possible energy costly strains and we may write the elastic energy in general form as
\begin{align} \label{eq:unimode_energy1}
    E = \frac{1}{2} \int \dee^2 \posr \left( \stiff_{1 1} \nonMechStrainOne^2 + \stiff_{2 2} \nonMechStrainTwo^2 + 2\stiff_{12} \nonMechStrainOne \nonMechStrainTwo\right) \, .
\end{align} 
in terms of these variables. The stiffnesses $\stiff_{ij}$ also define the constitutive relationship between the strains and the corresponding stresses, which is written compactly,
\begin{equation}\label{eq:constitutiveLinearMech}
    \begin{bmatrix}
        \mechStress \\ \nonMechStressOne \\ \nonMechStressTwo
    \end{bmatrix}
    = 
    \begin{bmatrix}
        0 &  0  & 0 \\
        0 & \stiff_{1 1} & \stiff_{1 2} \\
        0 & \stiff_{1 2} & \stiff_{2 2}
    \end{bmatrix}
    \begin{bmatrix}
        \mechStrain \\ \nonMechStrainOne \\ \nonMechStrainTwo
    \end{bmatrix} \, ,
\end{equation}
using the Voigt convention of treating stress and strain tensors as vectors of orthonormal components. Energy conservation requires the symmetry of this matrix, so that the presence of the mechanism eliminates three of the six independent stiffnesses. Thus, the defining property that the mechanism strain alone cannot generate stress also implies that mechanism stress with the same tensorial form cannot be supported $\mechStress = 0$.

It is always possible to choose coordinate axes such that the mechanism strain is proportional to a diagonal tensor:
\begin{align}\label{eq:mechStrain}
 \ttens{\hat{e}}_m
\equiv  \aA \begin{bmatrix} 1 & 0 \\ 0 & -\poisson  \end{bmatrix},
\end{align}
where the mechanism Poisson's ratio $\poisson$ is the negative of the ratio of the two principle strains and $\aA \equiv  \sqrt{2/(1 + \poisson^2)}$ is a normalization factor. Note that a quarter-turn of the coordinate system inverts the Poisson ratio, and therefore the mechanism strain can switch from auxetic ($\poisson<0$) to anauxetic $\poisson>0$ by passing through the   through uniaxial strains where the Poisson's ratio either vanishes or diverges.


Arbitrary mechanisms can be generated via a system of rigid rotating parallelograms joined at their corners by ideal frictionless hinges, as shown in the generalization of the canonical ``Rotating Squares'' structure shown in Fig.~\ref{fig:main1}a,b. Here, the mechanism motion is traversed by rotating the rigid parallelogram blocks in opposite fashion to their neighbors, around the ideal frictionless hinges that connect them. This generates finely detailed rearrangements within each unit cell, as well as an overall displacement of each unit cell's position (i.e. center of mass) which varies smoothly through space. The mechanism elasticity theory above then applies to a coarse description of the material deformations in terms of these smoothly varying unit cell positions, in which the mechanism strain (Eq.~\ref{eq:mechStrain}) describes the transformation of the lattice vectors (Fig.~\ref{fig:main1}a,b pink arrows) which link these unit cells to their neighbors. As shown in Fig.~\ref{fig:main1}d, varying the parallelogram angle $\designAngle$ and mechanism rotation $\mechAngle$ spans the possible values of the Poisson's ratio $\poisson$ and, paired with a rotation of the coordinate system, any desired linear mechanism strain may be probed.


\section{Sheared analytic modes \label{sec:shearedAnalytics}}

For the special case of pure-dilational mechanisms, zero-energy deformations are those that disallow shear and hence preserve angles in the material. As is well-known, maps with this property are complex-analytic. That is, when points in the real plane are mapped to the complex plane, $(z,z^{*}) \equiv (x + i y, x - i y)$ with the equivalent map  for displacements $(u,u^*) \equiv (u_x + i u_y, u_x - i u_y)$ (see, e.g.,~\cite{england2003complex}), the zero-energy (i.e. stress-free) deformations are precisely those that satisfy \emph{complex analyticity}, $\partial_{z^*} u = \partial_z u^* = 0$, which has yielded tremendous insight into dilational metamaterials~\cite{Sun2012, Czajkowski2022, zheng2021continuum}.

Thus motivated, we seek to extend this analyticity to generic mechanisms outside of the pure-dilational limit. 
A single transformation applied identically to both the material coordinates and the displacements cannot achieve this. However, this can be achieved via a pair of related but distinct transformations on the material coordinates and the displacements:
\begin{align}\label{eq:coordTransform}
    \begin{bmatrix} \ww \\ \bar{\ww} \end{bmatrix} &\equiv
    \begin{bmatrix} 1 & \frac{1}{\gamgam} \\ 1 &  -\frac{1}{\gamgam}  \end{bmatrix}
    \begin{bmatrix} x \\ y \end{bmatrix},
    \\
    \begin{bmatrix} u \\ \bar{u} \end{bmatrix} &\equiv 
    \begin{bmatrix} 1 & -\gamgam \\ 1 & \gamgam \end{bmatrix}
    \begin{bmatrix} u_x \\ u_y \end{bmatrix},
    \\
    \gamma &\equiv \frac{1}{\sqrt{\poisson}}.
\end{align}

\noindent
Note that $\gamma$ is real for anauxetic mechanisms and imaginary for auxetic ones. For dilational mechanisms, $\gamma = -i$ recovers the known conformal case in which displacements and material coordinates transform identically. These transformations also determine the gradients in the transformed coordinates, since we require  $\partial_\ww \bar{\ww} = \partial_{\bar{\ww}} \ww = 0$ and $\partial_\ww \ww = \partial_{\bar{\ww}} \bar{\ww} = 1$.

The utility of this transformation is seen in the identification of zero-energy deformations. As shown in Appendix~\ref{app:coordinateTransforms}, it transforms the requirement that the two non-mechanism strains vanish, into the requirement that two of the derivatives vanish:
\begin{align}\label{eq:softModeOne}
    \partial_{\bar{\ww}}u = 0 & \quad \rightarrow \quad u  = \softModeOne(\ww)  \\ \label{eq:softModeTwo}
    \partial_{\ww} \bar{u} = 0 & \quad \rightarrow \quad \bar{u}  = \softModeTwo(\bar{\ww})  \, .
\end{align}
Hence, for a continuum stress-free deformation, the transformed fields are each analytic functions of just one of the transformed coordinates.

In simply-connected domains, these functions may then be generated by  simple series expansions, e.g. $\softModeOne(\ww) = \sum_{n=0}^{\infty} C_n \ww^n$. The  requirement that $u_x, u_y$  be real-valued (simply taking the real value of arbitrary complex functions would violate analyticity) enforces nontrivial restrictions on the functions $\softModeOne, \softModeTwo$. On the auxetic side, where coordinates $\ww, \bar{\ww}$ are complex-valued, we require that $\softModeOne(\ww), \softModeTwo(\bar{\ww})$ be complex conjugates of one another. For the anauxetic side, $\ww, \bar{\ww}$ are real-valued and we simply require that $\softModeOne, \softModeTwo$ be real-valued functions. As this recipe for generating energy-free continuum unimode deformations relies on sheared analytic functions of a sheared coordinate system, we refer to them as ``sheared analytic modes''. Again, the exact mathematics of the conformal soft maps from~\cite{Czajkowski2022} is easily recovered in the limit $\poisson \rightarrow -1$.

In addition to such stress-free displacements, there are patterns of stress that satisfy the bulk equilibrium condition $\partial_i \sigma_{ij} = 0$~\cite{landau1986theory}. Upon making a similar transformation similar to Eq.~\ref{eq:coordTransform}
\begin{align}
    \label{eq:stress_vars}
    \begin{bmatrix} \stress \\ \bar{\stress} \end{bmatrix} &\equiv
    \begin{bmatrix} A & -\gamgam \\ A &  \gamgam \end{bmatrix}
    \begin{bmatrix} \nonMechStressOne \\ \nonMechStressTwo \end{bmatrix} \, ,
\end{align}
these conditions for force balance may be rewritten simply as $\partial_{\bar{\ww}} \stress = \partial_{\ww} \bar{\stress} = 0$. It is clear that this is again the same set of equations that were governing the stress-free displacement patterns. Beyond the immediate analytic insight this generates, Eq.~\ref{eq:stress_vars} indicates that the responses to ``force-free'' (i.e. stress-free) versus ``force-bearing'' loading are mathematically dual spaces; solutions to one problem can readily be transformed into a solution to a dual problem in the other. To confirm this suggestive result, we examine numerically force balanced states of our unimode material in both force-free (Fig.~\ref{fig:main2}a-d) and force-bearing (Fig.~\ref{fig:main2}e-h) load situations as the unit cell structure becomes finer. In both cases, following the analyses described in Appendix.~\ref{app:numerics}, our analytic framework captures the vast majority ($>95\%$) of observed deformation and stress, and the quality ubiquitously improves as the system size is increased. In other words, sheared analytic modes take hold and control response as the continuum limit of these materials is approached. Note that ``stress-free'' here refers to the deformations themselves while ``force-free'' refers to the loading conditions that generate them, and similarly for the ``stress-free'' patterns and loading conditions.

While the duality is most clearly illustrated in the stress variables of Eq.~\ref{eq:stress_vars}, the $\ww, \bar{\ww}$ formalism also allows the nontrivial integrations involved in determining force-balanced displacement fields from stresses to be surpassed. In Appendix~\ref{app:generalFBmodes} the general form for a force-balanced displacement field is derived, and is again composed purely from sheared analytic functions, which showcases the power of this formalism.


\section{Spatial character of generic unimode response \label{sec:spatialModes}}

\begin{figure}[t]  
\begin{center}
    \includegraphics[width=0.5\textwidth]{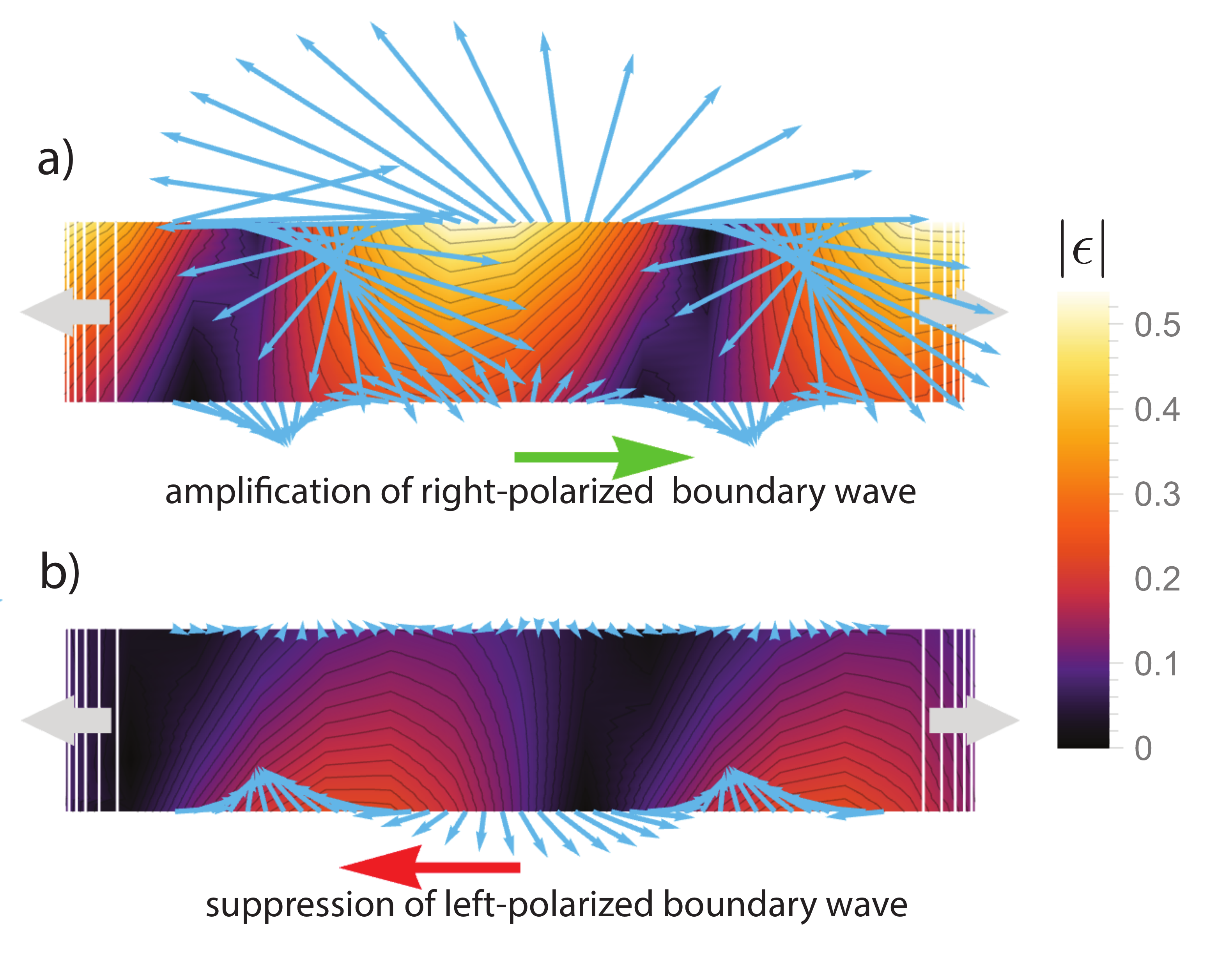}
    \caption{ \textbf{Auxetic Unimode for Filtering of Mechanical Signals} (\textbf{a}) For an auxetic mechanism, a long strip of material will amplify displacement loads with polarization to the right. Polarization of input signal is the direction of travel in which the displacements would appear to rotate counterclockwise. (\textbf{b}) Shows the opposite exponential suppression of the opposite polarization input. \label{fig:main4} }
    \end{center}
\end{figure}

As a mechanism is tuned from the auxetic to the anauxetic, the sheared coordinate systems $\ww$ and $\bar{\ww}$ briefly converge, becoming equal at $\poisson=0$ and then real for $\poisson>0$. This defines an exceptional point separating auxetic and anauxetic metamaterial mechanisms. Rather than controlling the more common energetic spectra~\cite{Miri2019} or phase transitions~\cite{Fruchart2021} for nonconservative systems, this is an exceptional point in a spatial coordinate transform, and it therefore distinguishes between spatial patterning types. On the auxetic side, the components of the displacement for a sheared analytic mode obey elliptic partial differential equations (see e.g.~\cite{stone2009mathematics}) and are harmonic conjugate functions of the sheared coordinates. On the anauxetic side, these components remain conjugate, but as real-valued functions outside of the complex analytic setting; these components obey partial differential equations of hyperbolic character.

To illustrate these response patterns, we consider the infinite half-plane, of arbitrary orientation, and with the component of displacement along the boundary fixed to an oscillatory function. As shown in Fig.~\ref{fig:main3}a,b,  the auxetic response decays into the bulk while simultaneously oscillating in a direction determined by both $\poisson$ and orientation of the mechanism principal axes. As $\poisson \rightarrow 0$ approaches the exceptional point from the negative side, the lengthscale of spatial decay diverges, eventually leading to persistent bulk oscillatory response in the anauxetic case. We note that, in this case the component of displacement along the edge is not sufficient to completely constrain the motion: the perpendicular displacement component at the boundary is free, and there is an infinite space of bulk modes which do not alter the boundary constraint which may be superimposed here, as described in greater detail in Appendix~\ref{app:halfPlane}. 

This behavior may be exploited in a long strip geometry. Here, an arbitrary displacement input on one boundary may be decomposed into two polarizations, which rotate in opposite directions along the boundary. As derived in Appendix~\ref{app:longStrip} and shown in Fig.~\ref{fig:main4}, one polarization decays exponentially into the bulk, while the other polarization (which was ruled out of the soft modes in the half-plane) will be amplified. As such, the unimode material acts not only as a mechanical amplifier but also as a filter that polarizes an initially generic static response. Note that, because of the duality, similar amplification and filtering will persist for stressed boundary conditions. Furthermore, as the structure traverses the mechanism motion, undergoing uniform large deformations, the Poisson's ratio itself changes sign as shown in Fig.~\ref{fig:main1}d. This filtering property may therefore be switched on and off, with uniform applied strains acting as a ``gating voltage'' for mechanical signal processing in analogy with transistors. As suggested by the phrasing, these properties may prove useful in mechanical circuitry and logic.


\section{Discussion \label{sec:discussion}}

We have shown that the presence of a single mechanism (unimode) in two-dimensional elasticity, as can be achieved in mechanical metamaterials, ubiquitously confines static response to dual spaces of stress-bearing and stress-free deformation. This spanning space of available response patterns is determined solely by the orientation and Poisson's ratio of the principal mechanism strain, thereby unifying response across microctructures and loading conditions. The tunable Poisson's ratio $\poisson$, and particularly the exceptional point at $\poisson = 0$, open the door to switchable elastic behavior which may become useful in metamaterial devices to amplify and filter signals in mechanical computing and circuitry. 

These properties seem to hinge on a type of mechanical criticality present in two-dimensional unimode materials. Mechanism strain and rotation constitute two fields whose spatial variation must satisfy the compatibility condition that the induced two-dimensional displacement vanish around a loop (Appendix~\ref{app:unimodeElasticity} and \cite{Czajkowski2021-2, Czajkowski2022, zheng2021continuum}). The two stress fields in bulk (away from the loading) likewise are constrained by the requirement that the two components of external force vanish. It is an open question, then, how such general principles extend to three-dimensional flexible mechanical metamaterials, either bulk ones or curved two-dimensional surfaces~\cite{Bertoldi2017, Schenk2013, Aharoni2014, Nassar2017, Nassar2022, Schenk2013, Overvelde2017, Griniasty2019}.

The duality between the spatial patterns of stresses and strains joins an impressive contingent of dualities in mechanics. Energy conservation and reciprocity imply a general conjugacy between uniform stresses and strains, Maxwell-Cremona dualities~\cite{Baker2013, Behringer2018} exist between force balance and position compatibility, and certain dilational metamaterials have recently been shown to possess self-dual phonon dispersions~\cite{Fruchart2020}, while general elasticity is itself dual to a tensor gauge theory~\cite{Pretko2018}.

\paragraph*{Acknowledgements}
We acknowledge helpful conversations with Corentin Coulais and Martin van Hecke.

\bibliography{bibliofile}



\onecolumngrid

\newpage

\appendix

\section{ \label{app:unimodeElasticity} Elasticity theory for generic planar mechanism metamaterials}

Here we expand on the argument in the main text, constructing the most general possible continuum theory governing elastic response in planar mechanism-based (i.e. unimode) metamaterials.  
In this section we develop a formalism to describe the coarse deformations of mechanism-based metamaterials. While an example class of mechanisms is given and analyzed as an example, the analytic approach will be generic and independent of the particular microscructure unless otherwise indicated. Further, we will focus on lattice mechanisms, meaning that the microstructure repeats periodically upon translation by either of a pair of lattice vectors $\latticeVecOne, \latticeVecTwo$. Working in the continuum-limit, where the microstructure of the unit cell becomes infinitesimally fine, means that strain gradient terms will be neglected in favor of the elastic energy terms coupling strains only.


\subsection{Mechanism motion in the continuum \label{app:continuumMechanism}}

Consider the mechanism depicted in main text Fig.~1a. The mechanism motion is traversed by rotating each light grey (rigid) quadrilateral by an angle $\mechAngle$, and simultaneously rotating the neighboring dark grey quadrilaterals by the opposite angle $-\mechAngle$ about the hinges at the corners where the blocks are attached. This motion, analogous to that of the canonical Rotating Squares lattice, will cost zero energy when the hinges are frictionless (i.e. perfectly flexible). More importantly, this motion generates a change in the lattice vectors. Undergoing a mechanism counter-rotation $\mechAngleInit \rightarrow \mechAngle$ changes these vectors from initial values $\latticeVecOneInit, \latticeVecTwoInit$ to final values $\latticeVecOne, \latticeVecTwo$. For a large finite system composed of $N_1, N_2$ unit cells in the respective lattice directions, the same uniform application of mechanism motion will change the macroscopic system shape according to $\{N_1 \latticeVecOneInit, N_2 \latticeVecTwoInit\} \rightarrow \{N_1 \latticeVecOne, N_2 \latticeVecTwo\}$. The change in the lattice vectors, as well as the change in the overall system shape are captured by a linear (affine) transformation $\latticeVec{i} = \ttens{\defTens}(\mechAngle) \cdot \latticeVecInit{i}$, where $i$ here indexes which lattice vector is being mapped, rather than the component. Note that this transformation tensor must depend implicitly on the choice of initial counter-rotation $\theta_0$ so that $\ttens{\defTens}(\mechAngleInit) = \ident$.

Any lattice mechanism may be parameterized in the manner of the previous paragraph and, for generality, we refer to a generic mechanism via a scalar $\mechScalar$ in place of $\theta$. In general, such a material may be described using coarse coordinates so that a unit cell initially located at $\posr$ will be located at $\posrFinal(\posr)$ after the mechanism motion. This is the approach of finite strain theory in the continuum. Here, $\posr$ captures a continuum of initial material coordinates, and is insensitive to the fine details of the lattice microstructure. The final continuum positions after deformation $\posrFinal(\posr)$, as well as the displacements $\posrFinal(\posr) - \posr$ are then smooth continuous vector fields defined over this reference position space $\posr$. The affine transformation $\ttens{\defTens}(\mechScalar)$, in this continuum context, is a uniform instance of a quantity known as the deformation gradient tensor. This tensor controls the transformation of the ``material infinitesimals'' which arbitrarily close points of the continuum material via $\dee\posrFinal = \ttens{\defTens}\cdot \dee \posr$. In this case of slowly varying continuum deformations, the lattice vectors take on the role of the material infinitesimals.

It is well-known in finite strain theory that the deformation gradient tensor may always be decomposed into $\ttens{\defTens} = \ttens{\rotationMat}\cdot \ttens{\rightStretch}$, where $\ttens{\rotationMat}$ is a rotation and $\ttens{\rightStretch}$ is a symmetric transformation known as the right stretch tensor. As we are considering a specific mechanism motion, rather than some generic deformation, $\ttens{\defTens}(\mechScalar) = \ttens{\rotationMat}(\coarseRot(\mechScalar))\cdot \ttens{\rightStretchMech}(\mechScalar)$ must be parametrized by the mechanism. Here, we have distinguished the right stretch tensor of the mechanism $\ttens{\rightStretchMech}(\mechScalar)$ via a particular tensor function, and $\coarseRot(\mechScalar)$ is the coarse rotation generated by the mechanism motion. As the mechanism motion is uniform, we may choose to rotate our coordinate frame along with the mechanism motion to eliminate $\coarseRot(\mechScalar)$ (i.e. we can always mix the mechanism with an energy-free rotation). It is clear, then, that tensor function $\ttens{\rightStretchMech}$ controls the strain of the metamaterial induced by the mechanism transformation. By a further choice of the orientation of the material reference (initial) coordinate frame, we may diagonalize the right stretch tensor induced by the mechanism $\ttens{\rightStretchMech}(\mechScalar) \rightarrow (( \principalOne(\mechScalar) , 0 ) , ( 0 , \principalTwo(\mechScalar)))$. These diagonal components are known as the principal stretches. While $\ttens{\rightStretchMech}$ will not stay diagonal for all mechanism final states $\mechScalar$, the parallelogram mechanisms explored in the main text have orthogonal lattice vectors which stay orthogonal all along the mechanism motion, and therefore the right stretch tensor also stays diagonal. The lattice vectors in this case may be written in compact form as $\latticeVecOne = 2 \cos(\designAngle + \mechAngle) \hat{x}$ and $\latticeVecTwo = 2\cos(\mechAngle) \hat{y}$. Then the principal stretches, given the starting mechanism point of $\mechAngleInit$, are $\principalOne(\mechAngle) = \cos(\designAngle + \mechAngle) / \cos(\designAngle + \mechAngleInit)$ and $\principalTwo(\mechAngle) = \cos(\mechAngle)/\cos(\mechAngleInit)$.

Note also that the mechanism right stretch is not an arbitrary tensor function, but must obey $\ttens{\rightStretchMech}(\mechScalarInit) = \ttens{\ident}$. Further, from knowledge of the right stretch induced by the mechanism, we may construct the metric of deformation (also known as the right Cauchy-Green deformation tensor) $\ttens{\metric}(\mechScalar) = \ttens{\defTens}^T \cdot \ttens{\defTens} \equiv \ttens{\rightStretchMech}(\mechScalar) \cdot \ttens{\rightStretchMech}(\mechScalar)$ and the Lagrange strain $ \ttens{\strain}^{(\text{Lagrange})}(\mechScalar)  \equiv \frac{1}{2}(\ttens{\metric}(\mechScalar) - \ttens{\ident})$ each generated by the mechanism motion.

The linear incrementation of the mechanism parameter by a small amount $\delta \mechScalar = \mechScalar - \mechScalarInit$ from an arbitrary starting state $\mechScalarInit$ will then induce a symmetric linear strain $\strain_{ij} = (\partial_i u_j + \partial_j u_i) /2$ of the coarse system. Again choosing to consider an oriented initial coordinate system which diagonalizes this generic strain, this may written in terms of the principal stretch functions as
\begin{align}
\begin{split}\label{eq:app_linearMechStrain}
  \ttens{\strain}(\delta \mechScalar) = & \delta \mechScalar \begin{bmatrix} \principalOne'(\mechScalarInit) & 0 \\ 0 & \principalTwo'(\mechScalarInit)\end{bmatrix} \\
   & = \delta \mechScalar  \principalOne'(\mechScalarInit)  \begin{bmatrix} 1 & 0 \\ 0 & -\poisson \end{bmatrix} \, ,
   \end{split}
\end{align}
where the primes denote derivatives with respect to the mechanism argument. Here, we have implicitly defined the mechanism-induced Poisson's ratio from the main text $\poisson \equiv -\principalTwo'(\mechScalarInit) / \principalOne'(\mechScalarInit)$. This corresponds to the conventional Poisson's ratio, which is defined from the ratio of the strain components $\strain_{11}$ and $\strain_{22}$, simplified using the knowledge that $\principalOne(\mechScalarInit) = \principalTwo(\mechScalarInit) = 1$. Again, for the example of the parallelograms this simplifies to $\poisson(\mechAngleInit) = -\tan(\mechAngleInit) / \tan(\designAngle + \mechAngleInit)$.

As an illustrative exercise, and to connect with the previous literature, we consider the possible spatial patterns of pure mechanism strain that are permitted in such metamaterials. While this setion concerns uniform applications of the mechanism thus far, the generalization to slowly varying strain patterns is natural, by promoting the continuum quantities to fields over the reference coordinates just like $\posrFinal(\posr)$. In particular, this is accomplished by establishing the mechanism field $\mechScalar(\posr)$ and coarse rotation field $\coarseRot(\posr)$ and demanding that the deformation tensor everywhere resemble $\ttens{\defTens}(\posr) = \ttens{\rotationMat}(\coarseRot(\posr))\cdot \ttens{\rightStretchMech}(\mechScalar(\posr))$. Of course, not any arbitrary patterns of $\mechScalar(\posr), \coarseRot(\posr)$ are possible: the spatial variation will generally create geometric inconsistencies. This problem is addressed by the well-known compatibility relations derived by demanding that any closed path integral of material infinitesimals be identically zero $\oint \dee \posrFinal = 0$ (i.e. this integral must vanish for every path in $\dee \posrFinal$ which corresponds to a closed loop in $\dee \posr$) . This leads to the both necessary and sufficient condition $\levi_{jk} \partial_k \defTens_{ij} = 0$ which must hold for $i=\{1,2\}$. Here, $\levi$ is the antisymmetric tensor with $\levi_{12} =1 $ Using the mechanism form of the deformation tensor, we arrive at a nonlinear first-order vector PDE
\begin{equation}\label{eq:app_finiteCompatibility}
  \vvec{\nabla} \coarseRot = \ttens{\specialCompatibilityMatrix}(\mechScalar)\cdot \vvec{\nabla} \mechScalar \, ,
\end{equation}
with
\begin{equation}\label{eq:app_specialBmatrix}
  \ttens{\specialCompatibilityMatrix}(\mechScalar) = \ttens{\levi} \cdot \ttens{\rightStretchMech}^{-1}(\mechScalar) \cdot \ttens{\levi} \cdot \ttens{\rightStretchMech}'(\mechScalar) \cdot \ttens{\levi} \, ,
\end{equation}
governing the soft mechanism strain patterns. As the left-hand side of this equation is independent of $\mechScalar$, we may integrate along a path from $\posr_{\text{init}}$ to $\posr_{\text{final}}$ and find a recipe to determine the rotation field from the mechanism field
\begin{equation}\label{eq:app_compatibilityIntegral}
   \coarseRot(\posr_{\text{final}}) - \coarseRot(\posr_{\text{init}}) = \int^{\posr_{\text{final}}}_{\posr_{\text{init}}} \dee\posr \cdot  \ttens{\specialCompatibilityMatrix}(\mechScalar)\cdot \vvec{\nabla} \mechScalar \, .
\end{equation}
For $\coarseRot$ to be a well-defined field, we will require that the integral in Eq.~\ref{eq:app_compatibilityIntegral} be path-independent. Similar to the derivation of the compatibility relation, this means that any integral in a loop must yield zero, and using Stokes' law we arrive at a necessary compatibility condition on the $\mechScalar$ field alone:
\begin{equation}\label{eq:app_finiteCompatibility2}
     0 = \levi_{ij} \partial_i \left[ \specialCompatibilityMatrix_{jk}(\mechScalar) \partial_k \mechScalar \right] \, .
\end{equation}
Further, as any field $\mechScalar$ satisfying Eq.~\ref{eq:app_finiteCompatibility2} may be integrated according to Eq.~\ref{eq:app_compatibilityIntegral} to define a compatible $\coarseRot$ field (up to an overall constant), this condition is also sufficient and either Eq.~\ref{eq:app_finiteCompatibility} or Eq.~\ref{eq:app_finiteCompatibility2} may be used to determine the compatible mechanism strain patterns.


\subsection{Generalized Voigt notation to describe mechanism elasticity \label{app:generalizedVoigt}}

Consider the standard procedure of constructing an energy functional for an elastic material. The deformation is captured by the smooth field of displacements $\vvec{u}(\posr)$, as defined in the previous section. Working in homogeneous space, without external fields, uniform displacements cannot incur any energy penalty, and we must turn to displacement gradients. In principle, the entire unsymmetrized strain $\unsymStrain_{ij}(\posr) = \partial_j u_i|_{\posr}$ is a candidate tensor field to control the energy density. However, in isotropic space there is another transformation in addition to the energy-free translations: the rotations. These rotations are mixed in with the energy costly strains in $\ttens{\unsymStrain}$. To account for this, the standard procedure is to notice that the rotation exclusively controls the anti-symmetric part of this tensor, while the symmetric part $\strain$ is independent of rotations. The energy functional is then built from this rotation-free part of the strain and may be generally written

\begin{equation}\label{eq:app_elasticEnergy}
  E = \frac{1}{2} \int \dee^2\posr \, \stress_{ij} \strain_{ij} \, ,
\end{equation}
where $\strain_{ij}$ and $\stress_{ij}$ are the familiar symmetric tensors capturing strain and stress, respectively. 

The mechanism strain identified in the previous section, similar to the rotation, cannot incur any energy, and must not appear in the elastic energy density. We are therefore motivated to identify the portions of strain which are orthogonal to this mechanism strain. To do this, we first must define a notion of orthogonality, by defining an inner product for symmetric tensors in two dimensions. Taking inspiration from the more general elastic energy Eq.~\ref{eq:app_elasticEnergy}, an appropriate inner product may be defined:
\begin{equation}\label{eq:app_innerProduct}
  \langle \ttens{T}_1 | \ttens{T}_2 \rangle \equiv \frac{1}{2} \mathrm{Tr}[(\ttens{T}_1)^\mathrm{T} \cdot \ttens{T}_2] \, .
\end{equation}
Here, the factor of $\frac{1}{2}$ reflects the number of dimensions we work in, so that the identity has unit norm. 

Given this inner product, there exists an infinite variety of orthonormal bases in which to break down stress and strain. Choosing, as in the previous section, to work in the coordinate system in which the linear mechanism strain is diagonal, our goal is to find a set of orthonormal unit tensors $\{ \hat{\ttens{\basisTens}} \}$, one of which is proportional to the mechanism strain. It is straightforward to check that this is satisfied by
\begin{align}\label{eq:app_tensorBasisMech}
  \basisTensMech & = \sqrt{\frac{2}{1 + \poisson^2}} \begin{bmatrix} 1 & 0 \\ 0 & -\poisson \end{bmatrix} \\ \label{eq:app_tensorBasisNonMechOne}
  \basisTensOne  & = \sqrt{\frac{2}{1 + \poisson^2}} \begin{bmatrix} \poisson & 0 \\ 0 & 1 \end{bmatrix}  \\ \label{eq:app_tensorBasisNonMechTwo}
  \basisTensTwo  & =  \begin{bmatrix} 0 & 1 \\ 1 & 0 \end{bmatrix}  \, .
\end{align}
As this constitutes a complete basis for symmetric tensors, we may break down the components of the strain $\strain_{ij} = \left[\mechStrain \basisTensMech + \nonMechStrainOne \basisTensOne + \nonMechStrainTwo \basisTensTwo\right]_{ij}$ and the stress $\stress_{ij} = \left[\mechStress \basisTensMech + \nonMechStressOne \basisTensOne + \nonMechStressTwo \basisTensTwo\right]_{ij}$ into the basis components.
For uniform systems, $\stress_{ij}$ is the derivative of the elastic energy with respect to $\strain_{ij}$, which generalizes to functional derivatives for non-uniform deformations.
This mechanism basis enables a notation which resembles that of Voigt, which is helpful for writing the constitutive relation:
\begin{equation}\label{eq:app_generalConstitutive}
  \begin{bmatrix} \mechStress \\ \nonMechStressOne \\ \nonMechStressTwo \end{bmatrix}
  =
  \begin{bmatrix} \stiff_{mm} & \stiff_{m1} & \stiff_{m2} \\ \stiff_{1m} & \stiff_{11} & \stiff_{12} \\ \stiff_{2m} & \stiff_{21} & \stiff_{22} \end{bmatrix}
  \begin{bmatrix} \mechStrain \\ \nonMechStrainOne \\ \nonMechStrainTwo \end{bmatrix}
  \end{equation}
In this notation, similar to that of Voigt, we must have a symmetric stiffness tensor $\stiff$ in order to connect to an equilibrium energy. Hence $\stiff_{1m} = \stiff_{m1}$ and $\stiff_{12} = \stiff_{21}$ and so on. While the notation here is chosen based on a mechanism, this relation is still completely general for a continuum elastic material in isotropic two-dimensional space. To account for the mechanism, we simply impose that the mechanism strain cannot generate any force nor energy nor stress. We then quickly may find that $\stiff_{mm} = \stiff_{m1} = \stiff_{m2} = 0$ and, with the tensor symmetry we have eliminated the row and column associated with the mechanism, leading to the main text constitutive relation. Further, this means that the mechanism stress $\mechStress$ will always be identically zero. Finally, this notation allows us to write the mechanism elastic energy via
\begin{align}\label{eq:app_elasticEnergy2}
  E = & \int \dee^2 \posr \langle \ttens{\stress} | \ttens{\strain} \rangle  \\ \nonumber
  & = \int \dee^2 \posr \left\{  \stiff_{11}\nonMechStrainOne^2 + \stiff_{22} \nonMechStrainTwo^2 + 2 \stiff_{12} \nonMechStrainOne \nonMechStrainTwo   \right\} \, .
\end{align}
Here, conventional strains $\epsilon_{ij}$ have a different normalization such that terms in the energy commonly appear with factors of $\frac{1}{2}$.

For completeness, the recipes to obtain these new strain variables from explicit spatial derivatives are 
\begin{align}\label{eq:app_mechStrainFromStrain}
  \mechStrain & = \frac{\aA}{2} (\strain_{11} - \poisson \strain_{22}) = \frac{\aA}{2}(\partial_x u_x - \poisson \partial_y u_y) \\ \label{eq:app_nonMechStrainOneFromStrain}
  \nonMechStrainOne & = \frac{\aA}{2} (\poisson \strain_{11} + \strain_{22}) = \frac{\aA}{2}( \poisson \partial_x u_x + \partial_y u_y) \\ \label{eq:app_nonMechStrainTwoFromStrain}
  \nonMechStrainTwo & = \strain_{12} = \frac{1}{2}(\partial_x u_y  +  \partial_y u_x)
  \, ,
\end{align}
where $\aA \equiv \sqrt{2/(1+\poisson^2)}$ is the normalization factor from Eqs.~\ref{eq:app_tensorBasisMech}~\&~\ref{eq:app_tensorBasisNonMechOne}. Note that a similar recipe may be used to obtain the mechanism vs. nonmechanism components of stress from $\stress_{ij}$. 
Finally, this tensor basis may still be useful in coordinate systems where the incremental mechanism motion is not a diagonal strain, simply by rotating the basis according to standard tensor transformation laws.


\section{Static response of continuum unimode structures \label{app:response}}

Given the elastic theory of Eq.~\ref{eq:app_elasticEnergy2}, we ask here what possible nonuniform response patterns are supported by this energy functional without generating a bulk force density. This is explored in the continuum limit, in the absence of strain gradient terms.


\subsection{Identifying useful coordinate transformations  \label{app:coordinateTransforms}}

Consider, for example, the pure dilational limit of a mechanism. We know from previous work that the standard transformation to complex coordinates $(z, z^*) = (x + i y, x - i y)$ along with $(u, u^* ) = ( u_x + i u_y, u_x - i u_y )$ greatly simplifies the analysis. In this case, the nonmechanism strains $\nonMechStrainOne, \nonMechStrainTwo $ (which are, in this case, the pure and the simple shears, respectively) become independent of the complex derivative fields $\partial_z u$ and $\partial_{z^*} u^*$. These are written
\begin{align} \label{eq:app_conformalShearOne}
  \nonMechStrainOne^{(\poisson \rightarrow -1)} & = \frac{1}{2}( \partial_z u^* + \partial_{z^*} u ) \\ \label{eq:app_conformalShearTwo}
  \nonMechStrainTwo^{(\poisson \rightarrow -1)} & = \frac{i}{2}( \partial_z u^* - \partial_{z^*} u )  
\end{align}
and we notice that choosing $u(z, z^*) \rightarrow u(z)$ captures the well known space of deformations which generate zero shear, known as the conformal maps. This is very convenient, as $u(z)$ may be expanded as a complex analytic function in powers of $z$, allowing this entire space of soft deformations to be generated by choosing a single list of coefficients. 

We therefore ask whether there might exist a coordinate transformation, similar to the complex plane transformation for pure dilational mechanisms, which simplifies the analysis of a more generic mechanism. Consider the generic candidate transformations
\begin{equation}\label{eq:app_dispTransform}
  \begin{bmatrix} \bar{u} \\ u  \end{bmatrix} = \begin{bmatrix} L_{11} & L_{12} \\ L_{21} & L_{22} \end{bmatrix}
  \begin{bmatrix} u_x \\ u_y  \end{bmatrix}
\end{equation}
and
\begin{equation}\label{eq:app_derivTransform}
  \begin{bmatrix} \partial_\ww \\ \partial_{\bar{\ww}} \end{bmatrix} = \begin{bmatrix} M_{11} & M_{12} \\ M_{21} & M_{22} \end{bmatrix} \begin{bmatrix} \partial_x \\ \partial_y  \end{bmatrix}
\end{equation}
where it will turn out to be more convenient to work with the transformation of the derivatives than the coordinates $\ww, \bar{\ww}$ directly.

To identify a useful transformation, we will attempt to identify which of the possible transformations will achieve the following goals:
\begin{enumerate}
  \item Wherever the condition $\partial_{\bar{\ww}} u = \partial_\ww \bar{u} = 0$ is satisfied, we will find the non-mechanism strains are zero $\nonMechStrainOne = \nonMechStrainTwo = 0$.
  \item Wherever the nonmechanism strains are zero $\nonMechStrainOne = \nonMechStrainTwo = 0$, we will find the partial derivatives  $\partial_{\bar{\ww}} u = \partial_\ww \bar{u} = 0$ also zero. 
  \item $u, \bar{u}$ should be produced from linearly independent transformations.
  \item $\ww, \bar{\ww}$ should be produced from linearly independent transformations.
\end{enumerate}
Searching for the conditions which allow the first two conditions to be simultaneously met, we arrive at a linear algebra problem
\begin{equation}
  \begin{bmatrix} 0 \\ 0 \\ 0 \\ 0 \end{bmatrix} =
  \begin{bmatrix}
    M_{11} & -\poisson M_{12} & 0 & 0 \\
    M_{12} & -M_{11} & 0 & 0 \\
    0 & 0 & M_{21} & -\poisson M_{22} \\
    0 & 0 & M_{22} & -M_{21}
  \end{bmatrix}
  \begin{bmatrix}
    L_{11} \\ L_{12} \\ L_{21} \\ L_{22}
  \end{bmatrix}
\end{equation}
which we have oriented as a problem to solve for the $L$ values given a predetermined set of $M$ values. This has separated into two problems, and in order to have non-trivial values of $L_{11}$ and $L_{12}$ we must have $\poisson M_{12}^2 - M_{11}^2 = 0$. Similarly, in order to not have nonzero values of $L_{21}$ and $L_{22}$ we require $\poisson M_{22}^2 - M_{21}^2 = 0$. Together, this leads to the collected conditions on our transformations
\begin{align} \label{eq:app_coordRestrictions}
  M_{12} & = s_1 \gamgam M_{11} \\ 
  M_{22} & = -s_1 \gamgam M_{21} \\ 
  L_{12} & = s_1 \gamgam L_{11} \\ \label{eq:app_coordRestrictionsFinal}
  L_{22} & = -s_1 \gamgam L_{21} \, ,
\end{align}
where the scalar $s_1$ may be chosen to be $+1$ or $-1$.
Any transformations satisfying these conditions will lend useful simplifications for a generic mechanism as explored in the next sections. This allows free nonzero choice of $M_{11}, M_{21}, L_{11}, L_{21}$ from which we may determine the rest of the constants. To connect with the conformal example, we make the simple choices $M_{11} = M_{21} = \frac{1}{2}$,  $ L_{11} = L_{21} = 1$ and $s_1=1$, leading to the transformations
\begin{align}\label{eq:app_coordinateTransforms}
  \partial_{\ww} & = \frac{1}{2} (\partial_x + \gamgam \partial_y) \\
  \partial_{\bar{\ww}} & = \frac{1}{2} (\partial_x - \gamgam \partial_y) \\
  \bar{u} & = (u_x + \gamgam u_y) \\ 
  u & = (u_x - \gamgam u_y) \, .
\end{align}
This also implies the corresponding direct transformation of the coordinate system
\begin{align}\label{eq:app_coordinateTransforms2}
  \ww & = x + \frac{y}{\gamgam} \\
  \bar{\ww} & = x - \frac{y}{\gamgam} \, ,
\end{align}
so that $\partial_{\bar{\ww}} \ww = \partial_\ww \bar{\ww} = 0$ and $\partial_\ww \ww = \partial_{\bar{\ww}} \bar{\ww} = 1$.
Writing our strain components in terms of these variables
\begin{align}\label{eq:app_mechStrainSpecialCoords}
  \mechStrain & = \frac{\aA}{4} (1 - \poisson^2)(\partial_\ww \bar{u} + \partial_{\bar{\ww}} u) +   \frac{\aA}{4} (1 + \poisson^2)(\partial_{\bar{\ww}} \bar{u} + \partial_\ww u)  \\ \label{eq:app_nonMechStrainOneSpecialCoords}
  \nonMechStrainOne & = \frac{\aA \poisson}{2} (\partial_\ww \bar{u} + \partial_{\bar{\ww}} u)  \\ \label{eq:app_nonMechStrainTwoSpecialCoords}
  \nonMechStrainTwo & = \frac{1}{2 \gamgam} (\partial_\ww \bar{u} - \partial_{\bar{\ww}} u)   \\ \label{eq:app_coarseRotSpecialCoords}
  \coarseRot & = \frac{1}{2 \gamgam} (\partial_{\bar{\ww}} \bar{u} + \partial_\ww u) \, ,
\end{align}
we may see immediately that the  space of deformations composed purely of mechanism strains $\mechStrain$ and rotations $\coarseRot$ (i.e. $\nonMechStrainOne = \nonMechStrainTwo = 0$) is identified with the conditions $\partial_\ww \bar{u}  =  \partial_{\bar{\ww}} u = 0$. Therefore, these are written most generally as analytic functions $u = \softModeOne(\ww)$ and $\bar{u} =  \softModeTwo(\bar{\ww})$, which span the space of continuum stress-free deformations that the unimode material can support, as described in the main text.

It is important to note that any coordinate transformation which satisfies the conditions Eqs.~\ref{eq:app_coordRestrictions}-\ref{eq:app_coordRestrictionsFinal} will become linearly dependent at the exceptional points $\poisson=0$ and $\poisson=\pm\infty$. At these points, the mechanism strain becomes uniaxial, and a one-dimensional coordinate system becomes the natural approach.




\subsection{Generating analytic force-balanced modes} \label{app:generalFBmodes}

We now investigate the possible patterns of response that arise under static loading. We aim beyond the stress-free analytic modes identified with the coordinate transformations of Sec~\ref{app:coordinateTransforms} to include those which may bear stress without generating bulk force. Continuing to employ these coordinate transforms, we may write the energy as
\begin{align}\nonumber
  E & = \frac{\gamgam}{2}  \int \dee\bar{\ww} \dee\ww \Big[  \frac{\stiff_{11} \aA^2 \poisson^2}{4} \left( \partial_w \bar{u}  + \partial_{\bar{w}} u \right)^2 \\ \label{eq:app_energy_wNotation}
    & + \frac{\stiff_{22} \poisson}{4} \left( \partial_w \bar{u} - \partial_{\bar{w}} u \right)^2 + \frac{\stiff_{12} \aA \poisson}{2 \gamgam} \left( (\partial_w \bar{u})^2 - (\partial_{\bar{w}} u)^2 \right) \Big].
\end{align}
We are searching for the force-balanced solutions to this energy, and therefore require that the functional derivatives with respect to the displacement fields, to which forces are proportional, vanish. While this is usually done with Cartesian components $u_x, u_y$, our notation allows the use of the $u, \bar{u}$ and $\ww, \bar{\ww}$.
Taking the functional derivative of this energy is then straightforward, and with a little algebra, the equations of force-balance are written
\begin{align}\label{eq:app_forcebalance1}
    0 & = \partial_\ww\left( B_1 \partial_\ww \bar{u} + B_2 \partial_{\bar{\ww}} u \right) \\ \label{eq:app_forceBalance2}
    0 & = \partial_{\bar{\ww}}\left( \bar{B}_1 \partial_{\bar{\ww}} u + B_2 \partial_\ww \bar{u} \right) 
\end{align}
where
\begin{align}\label{eq:app_the_bees}
    B_1 & \equiv \frac{\stiff_{11} \aA^2 \poisson^2}{2} + \frac{\stiff_{22} \poisson}{2} + \frac{\stiff_{12} \aA \poisson}{ \gamgam} \\
    \bar{B}_1 & \equiv \frac{\stiff_1 \aA^2 \poisson^2}{2} + \frac{\stiff_{22} \poisson}{2} - \frac{\stiff_{12} A \poisson}{ \gamgam} \\
    B_2 & \equiv \frac{\stiff_{11} \aA^2 \poisson^2}{2} - \frac{\stiff_{22} \poisson}{2}  \, .
\end{align}
Note that from Eqs.~\ref{eq:app_mechStrainSpecialCoords}-\ref{eq:app_coarseRotSpecialCoords} and the constitutive relation from Sec.~\ref{app:generalizedVoigt}, that this may be written quite simply in terms of stresses
\begin{align}\label{eq:forcebalanceStress1}
    0 & = \partial_\ww\left( \aA \nonMechStressOne +\gamgam \nonMechStressTwo \right) \\ \label{eq:app_forceBalanceStress2}
    0 & = \partial_{\bar{\ww}}\left(\aA \nonMechStressOne -\gamgam \nonMechStressTwo \right) 
\end{align}
and the natural definition of two stress scalars has become evident 
\begin{align}\label{eq:app_stressVar1}
    \stress & \equiv  \aA \nonMechStressOne - \gamgam \nonMechStressTwo = \stiffModeOne'(\ww) \\ \label{eq:app_stressVar2}
    \bar{\stress} & \equiv \aA \nonMechStressOne + \gamgam \nonMechStressTwo = \stiffModeTwo'(\bar{\ww}) \, ,
\end{align}
where we have made clear that the patterns of stress which obey force-balance will be again described by sheared analytic functions. Here, for convenience in the analysis that follows, we have chosen to write these functions $\stiffModeOne'(\ww), \stiffModeTwo'(\bar{\ww})$ as derivatives of analytic functions, which, due to the analyticity itself, does not sacrifice generality.

From knowledge of the stress patterns, the typical procedure is to use the constitutive relation to determine the strains, and then (non-trivially) to integrate these strains to obtain the displacement patterns. However, in this case the integration is facilitated by analyticity. Again using Eqs.~\ref{eq:app_stressVar1}\&\ref{eq:app_stressVar2} in the constitutive relations, we may solve for the displacement derivatives
\begin{align}\label{eq:app_solvingFBdisps1}
    \partial_{\bar{\ww}} u & = \frac{1}{B_1 \bar{B}_1 - B_2^2} \left( B_1 \stiffModeOne'(\ww) - B_2 \stiffModeTwo'(\bar{\ww}) \right) \\ \label{eq:app_solvingFBdisps2}
    \partial_{\ww} \bar{u} & = \frac{1}{B_1 \bar{B}_1 - B_2^2} \left( \bar{B}_1 \stiffModeTwo'(\bar{\ww}) - B_2 \stiffModeOne'(\ww) \right) \, .
\end{align}
These equations may then be integrated to obtain the general form of a force-balanced displacement:
\begin{align}\label{eq:app_FBdisps1}
    u & = \frac{1}{B_1 \bar{B}_1 - B_2^2} \left( B_1 \bar{\ww} \stiffModeOne'(\ww) - B_2 \stiffModeTwo(\bar{\ww}) \right) + \softModeOne(\ww) \\ \label{eq:app_FBdisps2}
    \bar{u} & = \frac{1}{B_1 \bar{B}_1 - B_2^2} \left( \bar{B}_1 \ww \stiffModeTwo'(\bar{\ww}) - B_2 \stiffModeOne(\ww) \right) + \softModeTwo(\bar{\ww}) \, , 
\end{align}
which conveniently distinguishes between the contributions which generate material stress $(\stiffModeOne, \stiffModeTwo)$ and those which do not $(\softModeOne, \softModeTwo)$.
It would seem at this point that the procedure is complete: given boundary conditions, the response pattern observed in the bulk should be described by the closed-form analytic solutions in Eqs.~\ref{eq:app_FBdisps1}~\&~\ref{eq:app_FBdisps2}. However, to generate an arbitrary member of the force-balanced modes, we must also acknowledge the requirement that the displacement components $u_x = \frac{1}{2}(u + \bar{u})$ and $u_y = \frac{1}{2 \gamgam} (\bar{u} - u)$ be real-valued. Recalling the definition of $\gamgam = \frac{1}{\sqrt{\poisson}}$, this requirement leads to different conditions on the functions $u, \bar{u}$ depending on the sign of $\poisson$, which determines whether $\ww, \bar{\ww}$ are real or complex-valued. When $\poisson>0$ and $\ww, \bar{\ww}$ are everywhere real-valued, the sheared analytic functions $\softModeOne, \softModeTwo, \stiffModeOne, \stiffModeTwo$ must simply be real-valued functions (i.e. the coefficients in the analytic expansion will be real) and may be chosen independently, yielding geometrically valid force balanced states. However, in the auxetic case  $\poisson<0$, $\ww$ and $\bar{\ww}$ are complex, and so are the nonuniform sheared analytic functions of these variables. The realness requirement of $u_x$ and $u_y$ then yields the nontrivial relations $\softModeOne(\ww) = \softModeTwo^*(\bar{\ww})$ and $\stiffModeOne(\ww) = \stiffModeTwo^*(\bar{\ww})$. While the functions are no longer independent, the dimension of the space of modes remains constant (the number of real scalars required to determine the soft mode is equal) because the modes are now defined via the real and imaginary parts of a single complex function rather than by two real functions.

The force-balanced modes in Eqs.~\ref{eq:app_FBdisps1}~\&~\ref{eq:app_FBdisps2} appear to diverge when $B_1 \bar{B}_1 - B_2^2 = (\stiff_{11}\stiff_{22} - \stiff_{12}^2) \aA^2 \poisson^3 \rightarrow 0$. This only occurs when either the reduced stiffness tensor has another zero eigenvalue, and hence the system becomes multimodal, or when $\poisson \rightarrow 0$. While the method presented here is not meant to handle the case of a bimodal metamaterial, it is important to note that the uniaxial case is still well-behaved (not actually divergent) when analyzed in terms of the real space displacements at $\poisson=0$.

Finally, we note that the duality of force-balanced stress and soft deformation is enhanced when considering the soft deformations in terms of strain rather than displacement. Similar to the nonlinear soft mode identification procedure identified in Sec.~\ref{app:continuumMechanism}, this is a question of solving the equation of mechanical compatibility of an unsymmetrized strain under the assumption of vanishing nonmechanism strain $\nonMechStrainOne, \nonMechStrainTwo  = 0$. Again, it is convenient to define transformed variables for these strains
\begin{align}\label{eq:app_strainTransform}
  \strainScalar & \equiv \aA \mechStrain - \gamgam \coarseRot \rightarrow \partial_w u \\
  \bar{\strainScalar} & \equiv \aA \mechStrain + \gamgam \coarseRot \rightarrow \partial_{\bar{\ww}} \bar{u} \, .
\end{align}
It is now evident that geometrically compatible soft strains will satisfy the equations $\partial_{\bar{\ww}} \strainScalar = \partial_\ww \bar{\strainScalar} = 0$. The duality of these soft deformations with the force balanced stress-bearing counterparts is even more direct in this notation.


\subsection{Force-balanced modes in the half-plane\label{app:halfPlane}}

Here we describe the specific spatial patterning of force-balanced deformation in the half-plane geometry. This is illustrative of the effect of an open boundary in more generic settings. In this geometry, the metamaterial sample extends infinitely far in the upper half-plane, away from an infinitely long open surface at $y=0$. For generality, we consider that the principal axes of the mechanism strain to be at some oblique angle $\interfaceAngle$ to our open boundary, and hence coordinate system. In this coordinate system the sheared analytic soft modes will be functions of the variables
\begin{align}\label{eq:app_rotW}
  \ww = & (\cos(\interfaceAngle) x + \sin(\interfaceAngle) y) + \frac{1}{\gamgam} (-\sin(\interfaceAngle) x + \cos(\interfaceAngle) y) \\ \label{eq:app_rotWbar}
  \bar{\ww} = & (\cos(\interfaceAngle) x + \sin(\interfaceAngle) y) - \frac{1}{\gamgam} (-\sin(\interfaceAngle) x + \cos(\interfaceAngle) y) \, .
\end{align}
The analytic functions constructed from these variables control the sheared displacements $u = \softModeOne(\ww)$ and $ \bar{u} = \softModeTwo(\bar{\ww})$, which are functions
\begin{align}\label{eq:app_rotU}
  u = & (\cos(\interfaceAngle) u_x + \sin(\interfaceAngle) u_y) - \gamgam (-\sin(\interfaceAngle) u_x + \cos(\interfaceAngle) u_y) \\ \label{eq:app_rotUbar}
  \bar{u} = & (\cos(\interfaceAngle) u_x + \sin(\interfaceAngle) u_y) + \gamgam (-\sin(\interfaceAngle) u_x + \cos(\interfaceAngle) u_y) \, ,
\end{align}
of the displacement components $u_x, u_y$ in this coordinate system.

In the half-plane, modes which exponentially grow or decay along the boundary will eventually become unphysical, due to the infinite nature of the boundary (such mechanisms will encounter nonlinear effects and eventually material failure at large strains). Therefore we search for soft modes which display no exponential growth along the boundary. Specifically, we will enforce the boundary condition in which the component of displacement pointing along the boundary follows an arbitrary oscillatory function $u_x(x,y=0) = u_x^0 \cos(q_0 x + \xi_0)$, while the transverse component of the boundary displacement will be left free.

Matching a sheared analytic mode to these boundary conditions depends on the geometry used. In the anauxetic case, we search for a solution of the form $u = \mathrm{Re}[u_0 \exp(i w q + i\xi)]$ and $\bar{u} = \mathrm{Re}[\bar{u}_0 \exp(i \bar{w} \bar{q} + i \bar{\xi})]$ (note that taking the real part does not violate analyticity for the anauxetic modes). To prevent unphysical exponential growth along the boundary we simply require here that $q$ and $\bar{q}$ be real numbers. Matching to the boundary condition, we find the following conditions:
\begin{align}
  \bar{u}_0 & = \frac{u_x^0}{(\frac{1}{\gamgam}\sin(\interfaceAngle) + \cos(\interfaceAngle))} \, , \\
  u_0 & = \frac{u_x^0}{( \cos(\interfaceAngle) -   \frac{1}{\gamgam}\sin(\interfaceAngle))} \, , \\
  q & = \frac{q_0}{(\frac{1}{\gamgam}\sin(\interfaceAngle) + \cos(\interfaceAngle))} \, , \\
  \bar{q} & = \frac{q_0}{( \cos(\interfaceAngle) -   \frac{1}{\gamgam}\sin(\interfaceAngle))} \, , \\
  \xi & = \bar{\xi} = \xi_0 \, .
\end{align}
This generates a bulk oscillatory mode as indicated in the main text.

However, we further note that any additional solution may be added to this, satisfying the above equations, after the trade $\bar{u}_0 \rightarrow -\bar{u}_0$, as this generates no impact on the boundary conditions. Further, the phase, amplitude, and wavenumber of this additional mode are free to be chosen to generate such a mode from the above conditions. Therefore, the boundary conditions on $u_x$ are not sufficient to fully constrain the mode. Here, oscillation still takes place along the directions along $\ww$ and along $\bar{\ww}$, independent of the orientation of the boundary, as shown in main text Fig.~3c,d.

In the auxetic case, we search for a solution of the form $u = u_0 \exp{[i w q]}$. To again prevent unphysical decay along the boundary, we require $q = \mathrm{Re}[q](1 +\frac{\tan(\interfaceAngle)}{\gamgam})$ which is now a complex number. The realness condition then determines the form of $\bar{u}(\bar{\ww}) = u^*(\ww)$ from complex conjugation; explicitly written as $\bar{u}(\bar{w}) = u_0^* \exp{[-i \bar{\ww}q^*]}$. Matching to the boundary condition, the soft mode takes the form
\begin{align}\nonumber
  u  = u_0 & \exp{\left[ i q_0 \left( x + \frac{(1-|\poisson|) \tan(\interfaceAngle)}{(1 + |\poisson| \tan^2(\interfaceAngle))} y\right)) \right]}  \\ \label{eq:app_auxeticHalfPlane}
  & \, \, \times \exp{\left[ -  |q_0| \sqrt{|\poisson|} \frac{1 + \tan^2(\interfaceAngle)}{1 + |\poisson| \tan^2(\interfaceAngle)} y \right]} \, .
\end{align}
Here, the absolute value of $q_0$ must be taken in the second (decaying/growing) exponential in order to ensure that the mode does not exponentially grow to unphysical quantities deep in the material bulk. While the amplitude and phase of $u_0$ must be determined to match precisely to the boundary condition, there is otherwise no freedom remaining and these boundary conditions are sufficient to completely determine soft response. The mode decays into the bulk with a characteristic length which diverges as $\poisson \rightarrow 0^-$ and simultaneously oscillates in the direction of the vector $(1, \frac{(1-|\poisson|) \tan(\interfaceAngle)}{1 + |\poisson| \tan^2(\interfaceAngle)})$.


\subsection{Force-balanced modes in the long strip\label{app:longStrip}}

To illuminate the possibility of a switchable mechanical filtering device, as suggested in the main text, we explore the geometry of the long strip. This is related to the half-plane geometry above, with a second infinite open boundary which is parallel and at a distance $d$ from the former, as shown in main text Fig.~4. In this case, the sheared analytic modes available to the anauxetic phase of the metamaterial are essentially unchanged. However, in the anauxetic phase, modes which were  divergent at long distances away from the open boundary due to exponential growth, may now be permitted. 

For an auxetic mechanism, any generic sheared analytic mode in this geometry may be composed of the exponential analytic functions $u(w) = u_0 \exp\left[ i q (1+\sqrt{|\poisson|} \tan(\interfaceAngle)) w \right]$, which are well-behaved (not divergent) along both boundaries. The complex conjugates of such functions will determine the $\bar{u}$ counterparts. We may therefore note that the contributions which have a positive value of $q$ will be exponentially suppressed by a factor $\exp\left[ - \frac{2 \pi \sqrt{|\poisson|}(1 + \tan^2(\interfaceAngle))}{\lambda (1 + |\poisson| \tan^2(\interfaceAngle))} d \right]$, where $\lambda$ is the wavelength of the mode at the bottom boundary.  Meanwhile, those with a negative value of $q$ will be amplified by the inverse of the same factor. These modes at the bottom boundary are differentiated by a polarization, in which one mode sweeps out a clockwise ellipse as $x$ is decreased (left polarized mode) while the other sweeps out a clockwise  ellipse as $x$ is increased (right polarized mode). It therefore follows that the effect of the metamaterial filter is to amplify the portions of an input mode with right-polarization, while suppressing those with left-polarization. The resulting signal at the far boundary of the strip will therefore be dominated by the right-polarized modes. This filtering behavior may be switched on and off by applying uniform mechanism strain to the system and tuning across the exceptional point of the system at $\poisson=0$.


\section{Numerical methods \label{app:numerics}}

\begin{figure}[t]  
\begin{center}
    \includegraphics[width=0.45\textwidth]{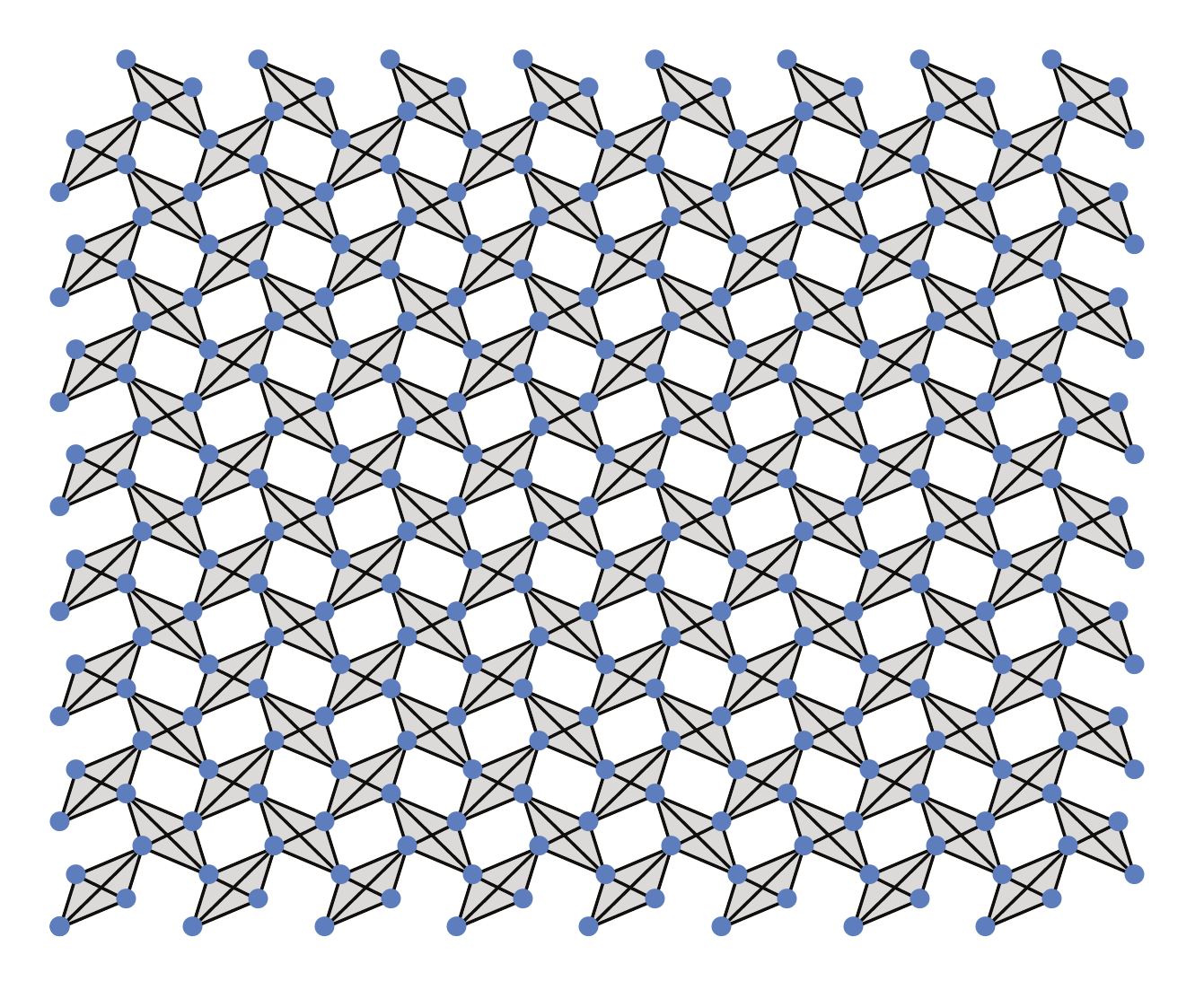}
    \caption{ \textbf{Diagrammatic representation of the constitutive model used to numerically obtain nonuniform force-balanced states of a mechanism.}  Blue dots represent the nodes, while black lines connect the node degrees of freedom with hookean springs. Grey parallelograms are for ease of viewing, representing the effective rigid quadrilateral elements of the mechanism in main text Fig.1. \label{fig:app_barSpring} }
    \end{center}
\end{figure}

Here, we describe the methods used to investigate force-balanced configurations of finite-sized mechanism lattices, as were used to generate data for main text Fig.~2. For simplicity, we employ lattices composed of identical unstretched ideal hookean springs, so that the energy is computed via $E = \frac{k}{2} \sum_\mu e_\mu^2$ where $e_\mu$ is the extension of bond $\mu$. As shown in Fig.~\ref{fig:app_barSpring}, the springs are placed to generate rigid parallelogram pieces joined at free hinges to emulate the mechanisms in main text Fig.~1: each rigid parallelogram is composed of 6 bonds and individually may only translate and rotate without energy cost. Because there is no energy penalty for pivoting the springs around the nodes they attach to, the mechanism in main text Fig.~1a,b is an energy-free motion of the lattice.

To generate force-balanced data in both the ``force-free'' and ``force-bearing'' conditions from the main text, we consider two distinct numerical processes listed below. In both cases, the finite mechanism metamaterial is defined via a system of identical hookean springs as depicted in Fig.~\ref{fig:app_barSpring}.

\subsection{Probing the stress-free continuum response in finite systems}\label{app:forceFree}

The loading patterns which are compatible with one of the sheared analytic modes $u = \softModeOne(\ww)$ and $\bar{u} = \softModeTwo(\bar{\ww})$ should lead to precisely such a zero-energy response in the continuum. This defines the ``force-free'' loading conditions. However, for finite systems with small but not infinitesimal unit cell size, these motions are not precisely zero-energy anymore. To examine the force-balanced low-energy configurations generated in a finite lattice, one may control the mechanism strain magnitude all along the left boundary of the system, as shown in main text Fig.~2a. To approximately control the mechanism strain in each of these boundary unit cells, a single stiff spring is added to the unit cell, crossing an open parallelogram void. These stiff springs are prescribed to extend according to a smoothly varying function through space. The springs are treated as rigid constraints, and the minimum energy state subject to these constraints is obtained. For linear mechanics, obtaining the force-balanced state of the system is reduced to a linear algebra problem which is readily solved in Mathematica. 

In the absence of nonuniformity, extending the boundary springs would trivially activate the mechanism alone. To probe the effect of nonuniformity, we choose the springs to extend according to a smoothly varying function through space. We would like to probe the trend in the effects of nonuniformity as the continuum limit is approached. We therefore probe these configurations as the system is made more dense with unit cells, keeping the smoothly varying function which controls the boundary conditions fixed. This defines our continuum limit procedure and, following the analyses in the following sections, the fitting to our analytic soft mode theory generically improves as this continuum limit is approached. 

For the data in the main text (Fig.~2a-d), the springs are extended according to a cubic function which is zero at the vertical midpoint and at top and bottom edges of the system. However, this choice is arbitrary and we have checked against various other functions which produce similar data.

\subsection{Probing the stress-bearing continuum response in finite systems}\label{app:forceBearing}

In contrast to the stress-free simulations designed above, we also investigate the stress-bearing configurations in the same system. To achieve this in a controlled manner, a new set of constraints are chosen to be generically incompatible with a sheared analytic mode. This defines the ``force-bearing'' loading conditions. To this end, rather than control the extension of additional bonds, we instead directly control the displacement of nodes all along the system boundary. To minimize mechanical boundary effects, we choose to prescribe the displacement of every node which has a different coordination than in the infinite lattice system (i.e. all dangling nodes). Such a set of nodes, along with the displacement constraints, are illustrated in main text Fig.~2e. The applied displacements along the boundary are determined by an arbitrary analytic force-balanced mode 
, known to generate finite stress even in the continuum limit. This is achieved using only the functions $\stiffModeTwo, \stiffModeOne$ defined in Eqs.~\ref{eq:app_FBdisps1}~\&~\ref{eq:app_FBdisps2}. 
For simplicity, only the second-order coefficients of the analytic expansions of these functions are explored here, with random phase and unit magnitude. More explicitly, we impose $\stiffModeOne(\bar{\ww}) \rightarrow \exp(i \eta_1)\bar{\ww}^2$ and  $\stiffModeTwo(\bar{\ww}) \rightarrow \exp(i \eta_2)\bar{\ww}^2$, with random numbers $\eta_1, \eta_2$. These simple forms are sufficient to generate nontrivial spatially varying stress patterns without calling into question the relative effects of different coefficients. However, including further terms in the analytic expansions of $\stiffModeOne, \stiffModeTwo$ is straightforward and generates similarly well-behaved data to that shown in the main text. 

To be compatible with the bulk force-balanced continuum mode, the fine motions within the unit cell must be accounted for. Rather than apply these smooth displacements directly to the nodes, force balance is first determined within each boundary unit cell, subject to the local strain experienced there. Together with the overall displacement of the unit cell itself, this determines the applied boundary conditions. These displacement conditions are again enforced using lagrange multipliers and the minimimum of the spring energy is identified via straightforward linear algebra. To observe the improving accuracy of our theory as the continuum limit is approached, the number of unit cells in each direction $N$ is increased to generate the data in main text Fig.~2g,h. 

Note that this particular procedure of detailed applied displacements was chosen to minimize boundary effects and observe the approach to continuum behavior in a controlled manner. However, it may be extended successfully to more realistic situations in which only one node is controlled per boundary unit cell, as well as the displacement along a partial boundary.


\subsection{Numerical estimation of strain and stress quantities from microscopic data}

\begin{figure}[t]  
\begin{center}
    \includegraphics[width=0.45\textwidth]{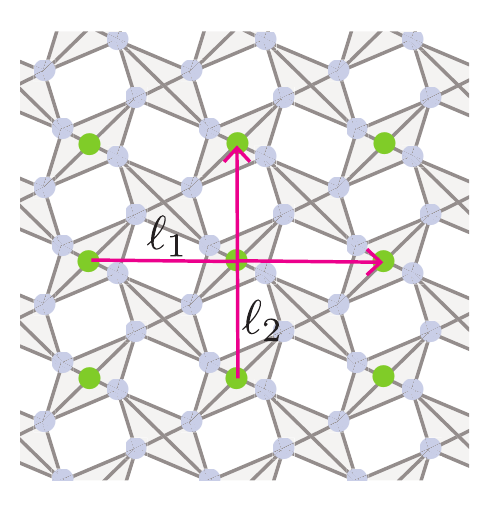}
    \caption{ Useful material vectors used in the estimation of local strain from deformed system configurations. Here, green dots represent the unit cell centers, and the strain for a single unit cell is most accurately obtained using the symmetric lattice vectors (pink arrows) which stretch between unit cell centers on opposite sides.  \label{fig:app_strainStressEstimates} }
    \end{center}
\end{figure}

From force-balanced configurations identified numerically in Secs.~\ref{app:forceFree}~\&~\ref{app:forceBearing}, we wish to estimate and separate the coarse mechanism strain from the nonmechanism strain and similarly separate the mechanism stress from the nonmechanism stress. To do this we must first construct methods to estimate conventional stress and strain in a coarse-grained approximation of the finite systems.

As discussed in the main text, the coarse strain locally controls the deformation of the lattice vectors. As shown in Fig.~\ref{fig:app_strainStressEstimates}a, we estimate the lattice vectors $\latticeVecOne, \latticeVecTwo$ after deformation to be those that connect the displaced center of mass of neighboring unit cells. Rather than constructing the symmetrized strain directly, we may estimate the unsymmetrized strain, i.e. the deformation gradient $\unsymStrain_{ij} \equiv \partial_j u_i$. In the continuum limit, this tensor controls the changes in lattice vectors $\Delta\latticeVec{i} \equiv \latticeVec{i} - \latticeVecInit{i} = \ttens{\unsymStrain}\cdot \latticeVecInit{j}$. The changes in two linearly independent lattice vectors are sufficient to estimate the unsymmetrized strain by constructing a matrix equation $(\Delta \latticeVec{1}, \Delta\latticeVec{2}) = \ttens{\unsymStrain} \cdot (\latticeVecInit{1} , \latticeVecInit{2})$, which is solved by
%
\begin{equation}\label{eq:app_strainInverse}
  \ttens{\unsymStrain}_{\text{est}} =
  \frac{1}{|\latticeVecInit{1} \otimes \latticeVecInit{2}|} \, \times \ttens{M} \cdot \ttens{\levi} \, ,
\end{equation}
where
\begin{equation}\label{eq:app_strainExtractionMatrix}
    M_{ij} = \left[(\Delta \latticeVec{2}\cdot \hat{\posr}_i) (\latticeVecInit{1}\cdot \hat{\posr}_j) - (\Delta \latticeVec{1}\cdot \hat{\posr}_i) (\latticeVecInit{2}\cdot \hat{\posr}_j) \right] \, ,
\end{equation}
$\levi$ is the antisymmetric unit tensor with $\levi_{12} = 1$, and $\{\hat{\posr}_i \} = ( \hat{\vvec{x}},\hat{\vvec{y}} )$ are the conventional unit vectors of the Cartesian coordinate system.
To reduce spurious finite-size effects in the estimation of strain, we employ an alternative definition of the lattice vectors which is more symmetric, as shown in Fig.~\ref{fig:app_strainStressEstimates}. It is straightforward to check that using these vectors, the errors in the strain estimation to first order in the lattice spacing $|\latticeVecInit{1}|$ are eliminated in favor of higher-order errors.

To then obtain the mechanism strain fraction, the numerically measured unsymmetrized strain is broken down into components
\begin{equation}\label{eq:app_mechStrainComponents}
  \ttens{\unsymStrain} = \mechStrain \basisTensMech + \nonMechStrainOne \basisTensOne + \nonMechStrainTwo \basisTensTwo - \coarseRot \ttens{\levi} \, ,
\end{equation}
using the orthonormal basis of tensors introduced in Eqs.\ref{eq:app_tensorBasisMech},\ref{eq:app_tensorBasisNonMechOne},\ref{eq:app_tensorBasisNonMechTwo}, 
and the antisymmetric tensor $\levi$ is the negative of the generator of rotations (and is also of unit norm). Using the inner product $\langle \basisTens_i | \basisTens_j \rangle = \mathrm{Tr}[\ttens{\basisTens}_i^T \cdot \ttens{\basisTens}_j]/2 = \delta_{ij}$, we obtain the strain amplitudes in this mechanism basis and are able to measure the nonmechanism strain fraction functional $\Delta_{\text{nonmech}}[\ttens{\strain}]$ plotted in main text Fig.~2c:
\begin{equation}\label{eq:app_nonmechstrainFraction}
  \Delta_{\text{nonmech}}[\ttens{\strain}] = \sqrt{\frac{\left\langle \nonMechStrainOne^2 + \nonMechStrainTwo^2 \right\rangle}{\left\langle \mechStrain^2 +  \nonMechStrainOne^2 + \nonMechStrainTwo^2 + \coarseRot^2 \right\rangle}} \, ,
\end{equation}
where $\langle \rangle$ is a spatial average taken over unit cells with a buffer region of two unit cells excluded at the boundary to minimize boundary effects.
This quantity thus compares the average magnitude of unsymmetrized strain to the magnitude of stress-bearing strain $\nonMechStrainOne, \nonMechStrainTwo$ in bulk soft deformations.

To estimate the local stress from the stress-bearing deformations employed in main text Fig.~2e-g, we employ the virial stress expression~\cite{Batchelor1970}. This standard formula approximates local mechanical stress from pairwise interactions.  The stress in unit cell $i$ located at $(x_i, y_i)$ is estimated using a symmetric lattice form
\begin{equation}\label{eq:app_virialStress}
  \ttens{\stress}(x_i, y_i) = \frac{k}{2 A_{\text{unit}}} \sum_{v \in i} \sum_{\mu \in v} \frac{k_\mu (|b_\mu| - b_\mu^0)}{|b_{\mu}|} \vvec{b}_\mu \otimes \vvec{b}_\mu
\end{equation}
where $A_{\text{unit}}$ is the unit cell area. Here the first sum is over the vertices $v$ contained in the unit cell while the second sum is over the bonds $\mu$ attached to the vertex $v$, with $\vvec{b_\mu}$ the deformed bond vector and $k$ the bond stiffness. With this symmetric summation convention, bonds stretching from one unit cell to the next contribute half of their stress to each unit cell while bonds between vertices within the unit cell are counted twice yielding the usual contribution.

Again, the local stress obtained in Eq.~\ref{eq:app_virialStress} may be broken into mechanism and nonmechanism pieces (there is no antisymmetric rotational piece due to the manifest symmetry of the stress tensor). The mechanism stress fraction functional $\Delta_{mech}[\ttens{\stress}]$ is then evaluated using
\begin{equation}
  \Delta_{\text{mech}}[\ttens{\stress}] = \sqrt{\frac{\left\langle \mechStress^2 \right\rangle}{\left\langle \mechStress^2 +  \nonMechStressOne^2 + \nonMechStressTwo^2\right\rangle}} \, .
\end{equation}

\subsection{Analytic fitting of sheared analytic modes and force-balanced stresses}

Finally, to estimate the validity of the analytic theory describing the global character of the force-balanced deformations, we search for the closest sheared analytic mode which fits with the data. This is accomplished by minimizing the fitting error
\begin{equation}\label{eq:app_fittingerror}
\delta_{\text{fit}}[\vvec{u}(), \{ u_k \}] = \sum^{N_p}_k |\vvec{u}(\posr_k) - \vvec{u_k}|^2 \, ,
\end{equation}
where $\vvec{u}(\posr_k)$ is vector displacement field  generated by the candidate analytic mode at the point $\posr_k$, and the sum is taken over the $N_p$ bulk material data points $k$. Again, a boundary layer of thickness two unit cells is excluded to minimize boundary effects. The candidate analytic mode is determined by a set of scalar coeffients. We choose to cut off these coefficients at a number $N_c = \text{min}(N_p/3, 20)$ to avoid overfitting. In the anauxetic case, there are then $N_c$ real scalar numbers determining the function $u = \sum_n C_n \ww^n$ and another $N_c$ independent real scalar numbers determining the function $\bar{u} = \sum_n D_n \bar{\ww}^n$. In the auxetic case, there are only $N_c$ coefficients determining $u$, which then determines the function $\bar{u}$. However, these coefficients are complex-valued, and the information required to define each function remains the same. The error in Eq.~\ref{eq:app_fittingerror} is minimized over these coefficients, which is a straightforward linear algebra problem. Estimating the error of these fits for main text Fig.~2d is accomplished by the fractional error $\Delta_{\text{fit}}[u, \{ u_k \}]$ (i.e. the square root of the fraction of variance unexplained)
\begin{equation}\label{eq:app_fittingFractionalError}
    \Delta_{\text{fit}}[\vvec{u}(), \{ u_k \}] = \sqrt{\frac{\delta_{\text{fit}}[\vvec{u}(), \{ u_k \}]}{\delta_{\text{fit}}[\vvec{u}(), \{ u_k \rightarrow 0 \}]}} \, ,
\end{equation}
where the term in the denominator gives the square magnitude of the displacement itself.
Identifying the closest analytic fit for the dual strain patterns, and therefore the fractional error one may utilize the exact process above, except with the replacements $u_x \rightarrow \aA \nonMechStressOne$ and $u_y \rightarrow \nonMechStressTwo$. Analogous fitting and error estimation methods have been used in related previous work~\cite{Czajkowski2022}.

\end{document}